\documentclass[draft]{agujournal}
\draftfalse

\journalname{ArXiV}

\begin{document}

\title{Thermodynamic constraints on the size distributions of tropical clouds}

\authors{Timothy J. Garrett, Ian B. Glenn, Steven K. Krueger}

\affiliation{1}{Department of Atmospheric Sciences, University of Utah, 135 S 1460 E, Rm 819, Salt Lake City, Utah, 84112}
\correspondingauthor{T.J. Garrett} {tim.garrett@utah.edu} 

\begin{keypoints}
\item Cloud perimeter distributions follow power-laws along moist isentropes and negative exponentials across moist isentropes
\item  The total perimeter of all clouds scales as the square-root of the atmospheric moist static stability
\item Thermodynamic arguments yield statistics for cloud ensembles that closely agree with detailed deterministic models.
\end{keypoints}


\begin{abstract}
Tropical convective clouds evolve over a wide range of temporal and spatial scales, and this makes them difficult to simulate numerically. Here, we propose that their statistical properties can be derived within a simplified time-independent co-ordinate system of cloud number $n$, saturated static energy $h^\star$, and cloud perimeter $\lambda$. Under the constraint that circulations around cloud edge compete for buoyant energy and air, we show that the product of cloud number and cloud perimeter $n\lambda$ is invariant with $\lambda$ and that cloud number follows a negative exponential with respect to cloud-edge deviations of $h^\star$ from the mean. Overall, the summed perimeter of all clouds scales as the square root of the atmospheric static stability, which suggests that the complexity of cloud field structures can be viewed statistically as an emergent property of atmospheric bulk thermodynamics.  Analytically derived conclusions are compared with a detailed tropical cloud field simulation and found generally to agree to within $\leq$13\%. For the sake of developing hypotheses about cloud temporal evolution that are testable in high resolution simulations, the shapes of tropical cloud perimeter distributions are predicted to be invariant as climate warms, although with a modest increase in total cloud amount.
\end{abstract}

\section{Introduction}

Despite rapid advances in computing speed, cloud modeling has proved to be a particularly stubborn problem for climate studies. Clouds evolve quickly and over a very wide range of spatial scales \citep{Stephens2005,Bony2006,IPCC_WG12007}. Turbulent eddies of air cascade from kilometers down to millimeter spatial scales. Microphysical processes are orders of magnitude
smaller and faster again. It has not been possible
to explicitly model all these details without shrinking model domains to climatologically irrelevant scales \citep{Krueger1997}.
 
Independent of model resolution, small-scale model behaviors are subject to larger-scale thermodynamic constraints. Taken as a whole, the atmosphere has an average density of air and rate of energetic throughput. The long-term stability of the temperature profile is determined by a balance between vertical gradients in radiative energy deposition and moist convective circulations of air. Internal variability of the system must satisfy this mean state, in which case individual components of the atmosphere compete with others for available energy and air.

As an example, it has previously been shown that convective mass fluxes within individual clouds have frequency distributions that follow negative exponentials. To obtain this result, what was assumed was only that the cloud ensemble could be characterized by a mean mass flux and that the associated timescales for convection were sufficiently rapid that individual clouds could be assumed to be quasi-independent  \citep{CraigCohen2006}. 

What has yet to be explored from this type perspective is how air is exchanged laterally in circulations between convective clouds and clear skies. This paper approaches the problem by considering a moist tropical region over the oceans containing a field of clouds that is in steady-state with respect to the larger-scale environment.  The focus is on the edges of the clouds, where there is an interface between saturated and unsaturated air across which contrasts in buoyancy and saturated static energy drive an evolving turbulent exchange. Looking at the dynamics of these exchanges, we derive number distributions for cloud perimeter as a function of larger scale atmospheric thermodynamic properties, and then compare these analytical expressions with results from a high-resolution simulation of an oceanic tropical cloud field.    
 
\section{Numerical simulation of convection}
In a convectively unstable atmosphere over tropical oceans, atmospheric circulations develop from the temporary instability created by surface solar heating and tropospheric radiative cooling, which forces the atmospheric lapse rate away from radiative-convective equilibrium. Potential energy builds and is then dissipated through work done by  clear and cloudy sky atmospheric convection \citep{Renno1996}. Where clouds form, they dominate the upward flow of air because efficient latent heat release during condensation produces buoyancy \citep{CraigCohen2006}. 

At the same time, clouds are leaky conduits for this vertical buoyant energy transfer because they generate turbulence; in a rising plume, cloudy air is exchanged with its stable subsaturated environment through lateral entrainment and detrainment. Detrainment tends to occur at the same level or higher than the level of entrainment \citep{Heus2008}; an overall radiative-convective equilibrium is reestablished because air with high buoyancy is entrained near cloud base and detrained near cloud top while negatively buoyant air subsides in surrounding clear skies.

\begin{figure}
\centering
\includegraphics[width=0.95\linewidth]{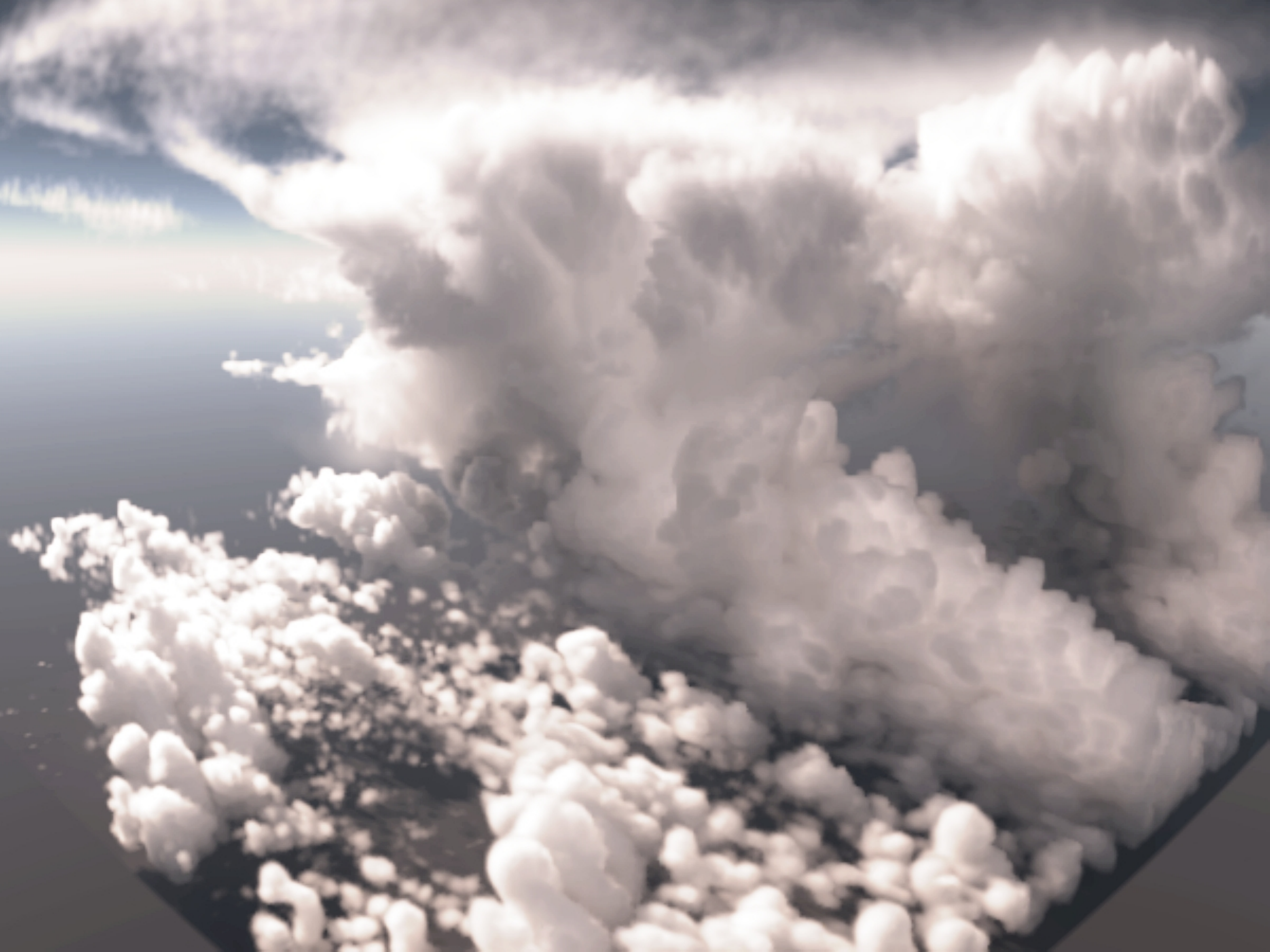}

\caption{\label{SHDOM} Visualization of the cloud condensate mixing ratios $q_c$ in the Giga-LES high-resolution large eddy simulation \citep{Khairoutdinov2009}. The volume shown is 20 km to a side and about 15 km tall, about one one-hundredth of the full Giga-LES simulation domain. A radiative transfer model \citep{Evans1998} is used to simulate the interactions with visible light.}

\end{figure}

Numerical models can be used to reproduce these dynamics by using a three dimensional spatial array of grid cells and equations that represent the flows of air that arise from buoyancy and horizontal pressure gradients. At a minimum, the state of a cell with volume $V$ at any given location and time is defined by a mass of air $m$ with density $\rho=m/V$, and mixing ratios for water vapor $q$ and cloud condensate $q_c$. Thermodynamic properties are introduced through the moist static energy $h$:
\begin{linenomath*}\begin{equation}
h =  c_{p}T+gz+Lq\label{eq:h}
\end{equation}\end{linenomath*}
where $c_{p}$ is the specific heat of dry air at constant pressure, $T$ is the air temperature, $g$ is the gravitational acceleration, $z$ height, and $L$ is the latent heat of water.  Thus, $h$ can be considered as the sum of potentials for molecular translational, rotational, and expansion motions,  ($c_p T$), falling ($gz$) and condensation ($Lq$). 

Within clouds, it is often assumed that water vapor is saturated with $q = q^\star\left(T,p\right)$, so that $h\left(q = q^\star\right) = h^\star$ is the saturated static energy. Constant $h^\star$ surfaces are often termed moist isentropes since, at constant pressure $p$, $d{s^\star} = d{h^\star}/T$ where $s^\star$ is the specific moist entropy; that is, in the absence of net diabatic heating, $h^\star$ and $s^\star$ are constant.  In terms of the equivalent potential temperature at saturation $\theta_e^\star$, the relationships are $ds^\star = c_pd\ln\theta_e^\star$ or $dh^\star = c_p{T}d\ln\theta_e^\star$. Outside clouds, where air descends along dry isentropes, the conserved variable is $h^0 = h\left(q = 0 \right)$.

One of the more useful numerical simulations for representing the evolution of these state variables is the ``Giga-LES'', which has served as a high resolution large eddy simulation (LES) benchmark for tropical cloud field evolution \citep{Khairoutdinov2009}. As a brief summary, the Giga-LES is initialized with idealized profiles from the GATE Phase III campaign, and it solves the anelastic system of momentum equations at 2\,s timesteps for a 24 hour period in a domain 204.8 km $\times$ 204.8 km $\times$ 19 km. Grid spacing is set to 100 m horizontally and 50 m vertically below 1 km height, increasing to 100 m by 5 km height for a total of $2048 \times 2048 \times 256 = 1,073,741,824$ grid points.  Model physics includes fluid dynamics, radiation, precipitation, cloud water, and sub-grid scale turbulence closure. A steady cooling profile of about 5 K d$^{-1}$, moistening of 1 g kg$^{-1}$ d$^{-1}$, and small surface temperature perturbations initiate a field of small cumulus clouds by the fourth hour that grow into several large cumulonimbus clouds by the eighth hour. Convective quasi-equilibrium occurs by the twelfth hour and continues for the rest of the 24-hour simulation. Vertical velocity statistics of the simulated clouds in the second twelve hours compare well with aircraft measurements taken during the GATE Phase III campaign \citep{Khairoutdinov2009, LeMoneZipser1980}. 

An illustration of the high resolution complexity and realism of the Giga-LES simulations is Figure \ref{SHDOM}, a 3D radiative transfer visualization of just 1\% of the full model domain. Despite the apparent realism of the simulation at 100 m resolution, it inevitably still misses much that is important. Even if spatial resolution were increased by orders of magnitude, small-scale turbulent and micrometer scale microphysical interactions would remain to be parameterized or ignored. Model runs are already computationally expensive: the simulated 24 hours of the Giga-LES required about 300,000 processor hours using the IBM Blue Gene/L "New York Blue" supercomputer at the New York Center for Computational Sciences, and an estimated $10^{18}$ FLOPs (M. Khairoutdinov, pers. comm.). This begs the question of whether there might be an alternative approach to increasing model complexity for increasing accurate representations of cloud ensembles.


\section{Saturated static energy as a point of neutral buoyancy at the cloud perimeter}\label{sec:MSE differences}


\begin{figure}
\centering
\includegraphics[width=11.4 cm]{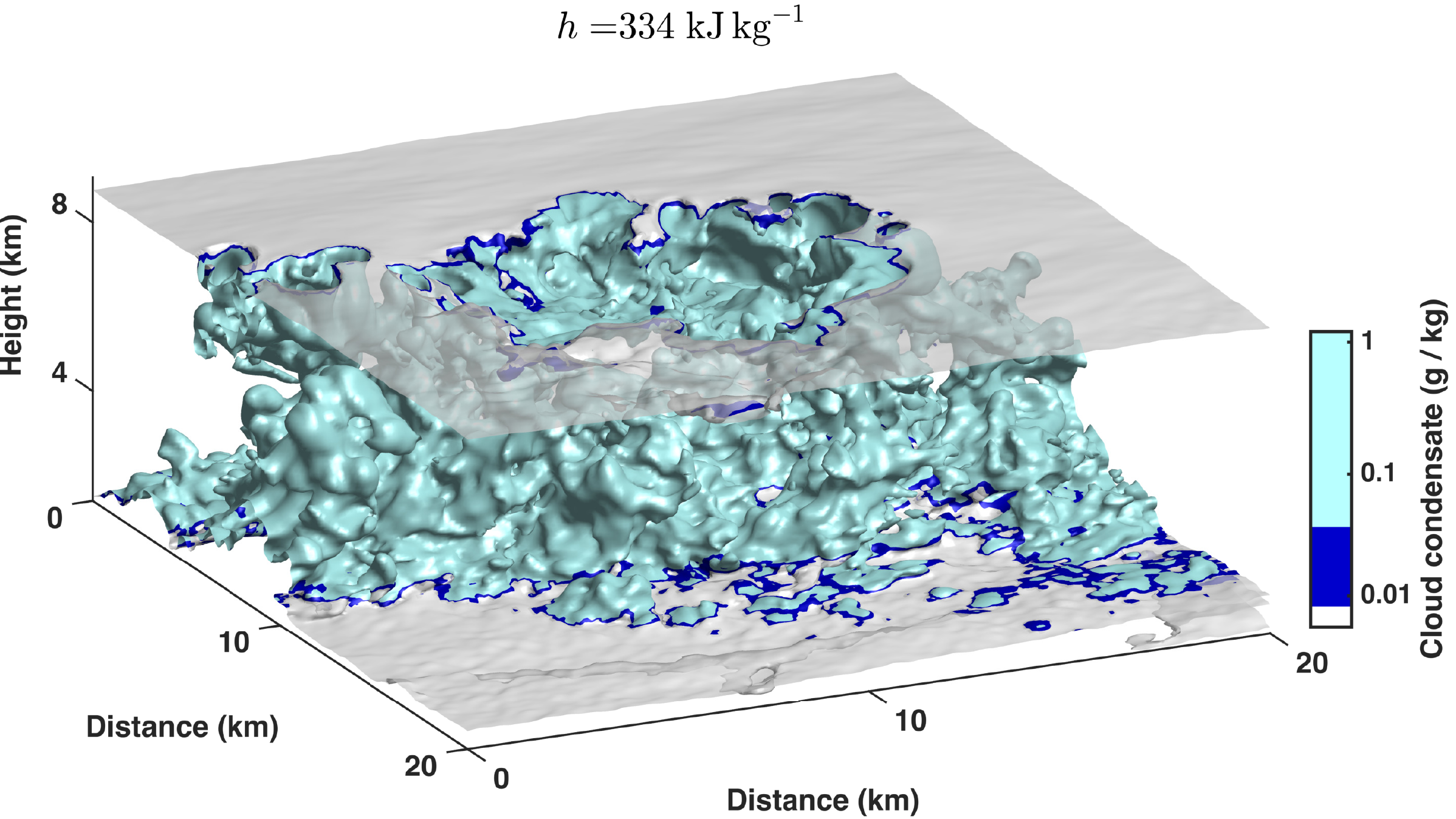}
\caption{\label{MSE_ISO_and_PDF} 334\,kJ\,kg$^{-1}$ isosurface of $h$ at a single time-step in a subset of the full Giga-LES domain where cloudy values of $q_c$ along that surface are shown in cyan and clear air values in grey. The dark blue line represents the subset of the cloud where the cloud condensate mixing ratio just crosses saturation with values of $q_c$ between 0.01 g/kg and 0.04 g/kg. Cloud perimeters along the $h$ isosurface are calculated as the total length of the closed contours, including both top and bottom. }
\end{figure}

\begin{figure}
\centering
\includegraphics[width=0.5\linewidth]{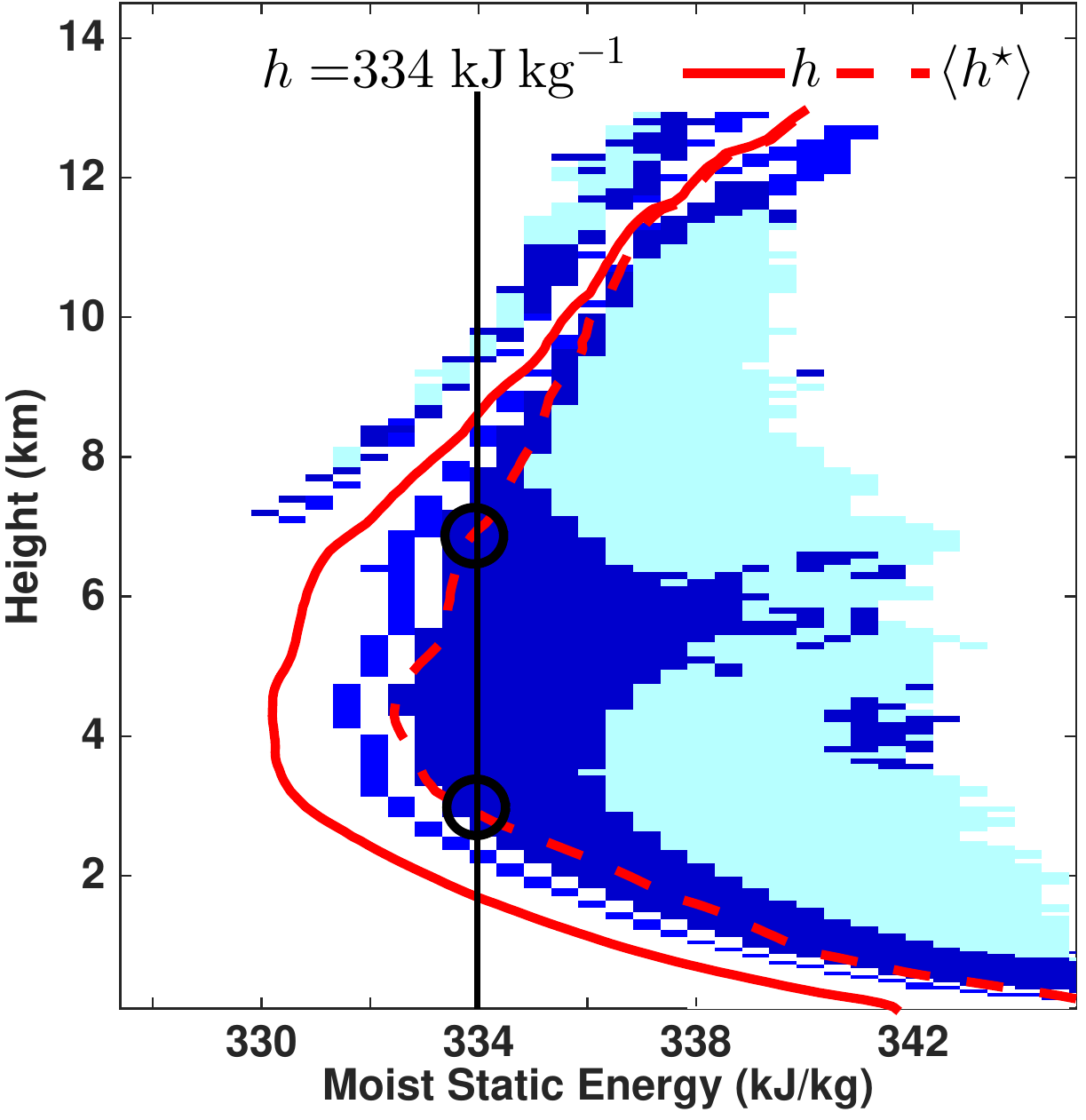}
\caption{\label{Isosurface} Horizontally-averaged values of $h$ in cloud for the full domain at the time step shown in Figure \ref{MSE_ISO_and_PDF}. Applying the same color scheme, cloud edge values with  mixing ratio values between 0.01\,g\,kg$^{-1}$ and 0.04\,g\,kg$^{-1}$ are shown in dark blue, and cloudy values with $q_c > 0.04$\,g\,kg$^{-1}$ in cyan. The domain-averaged moist static energy $h$ and saturated static energy $\left<h^\star\right>$ at any given level are shown in red. The black line corresponds to a 334\,kJ\,kg$^{-1}$ isosurface of $h$.}
\end{figure}

A possible pathway towards simplification is to develop a revised formulation for contrasts between clouds and clear skies, by shifting from a co-ordinate system expressed in space and time to something more explicitly thermodynamic. Currently, numerical models such as the Giga-LES precisely represent cloud boundaries as a temporally and spatially complex discretized surface where the value of $q_c$ crosses from zero to some very small value of $q_c$ (e.g. 0.01 g/kg). As an alternative, this boundary could be represented in terms of a co-ordinate system of  $h^\star$ for an ensemble of clouds in number $n$ and perimeter $\lambda$, based on the intuitive premise that thermodynamic exchanges of moist static energy and air between an ensemble of clouds and clear-skies are across an interface between the two, a location where air is just saturated. 

Figure \ref{MSE_ISO_and_PDF} shows intersecting isosurfaces of $q_c = 0.01$\,g\,kg$^{-1}$ and $h = 334$\,kJ\,kg$^{-1}$ for one section of the Giga-LES domain dominated by a single cloud. Mathematically, the intersection of any two surfaces is a line, here represented by the perimeter of the cloud where air is just saturated and $h\simeq h^\star$. The $h$ isosurface is very roughly a hyperboloid, intersecting the cloud perimeter at two distinct heights.  What is implied is a lateral gradient in $h$ between the interior and the exterior of the cloud normal to the cloud perimeter. 

For the Giga-LES cloud ensemble, Figure \ref{Isosurface} shows that cloud edge values of $h$ are closely aligned with the domain mean value of the saturated static energy at that height $\left<h^\star\left(z\right) \right>$. Averaged over clear and cloudy air, the atmosphere is sub-saturated. So, $\left<h^\star\right>$ represents the value of $\left<h\right>$ that would be obtained if the ensemble mean were moistened to the point that $\left<q\right>=\left<q^\star\right>$. Thus, for a partly cloud atmosphere, $\left<h\right> \leq \left<h^\star\right>$ at any given level, and cloudy air and clear air have greater and lesser values of $h$, respectively. 

Given the average relative motions of cloudy and clear air, what this suggests is that $\left<h^\star\right>$ serves as a point of neutral buoyancy with respect to cloud edge, with higher and lower values of $h$ and buoyancy on either side. Thermodynamically speaking, at a given height, $h^\star$ varies only as a function of temperature because $q^\star$ is only a function of temperature (Eq. \ref{eq:h}). On this basis, \citet{Randall1980} showed that if perturbations in total water $q_{tot} = q + q_c$ are small, then buoyancy perturbations scale linearly with perturbations in $h$ in dry air, and with perturbations in $h^\star$ in saturated air. This point was emphasized in numerical studies by \citep{Heus2008subsidingshells} and  \citet{GlennKrueger2014}. In active convection, buoyant air on the cloudy side of cloud edge can be characterized by $h = h^\star>\left<h^\star \right>$, whereas clouds are surrounded by narrow shells of clear air with negative buoyancy and values of $h<\left<h^\star \right>$ due to evaporative cooling. 



\section{Energetics of circulations around cloud perimeter}

Thus, on average, it seems that cloud edges can be seen as a point of neutral buoyancy, with departures from neutral buoyancy on either side that can be quantified in terms of differences in moist static energy with respect to the domain mean saturated static energy at that height. This is useful because we can now consider how these differences relate to the bulk atmospheric humidity and stability, and in turn to circulations of air around the neutral buoyancy point. Air rises and sinks, so through continuity it is entrained and detrained around cloud edges in buoyant circulations that mix cloudy air with clear air \citep{RaymondBlythe1986}. 

\begin{figure}
\centering
\includegraphics[width=0.95\linewidth]{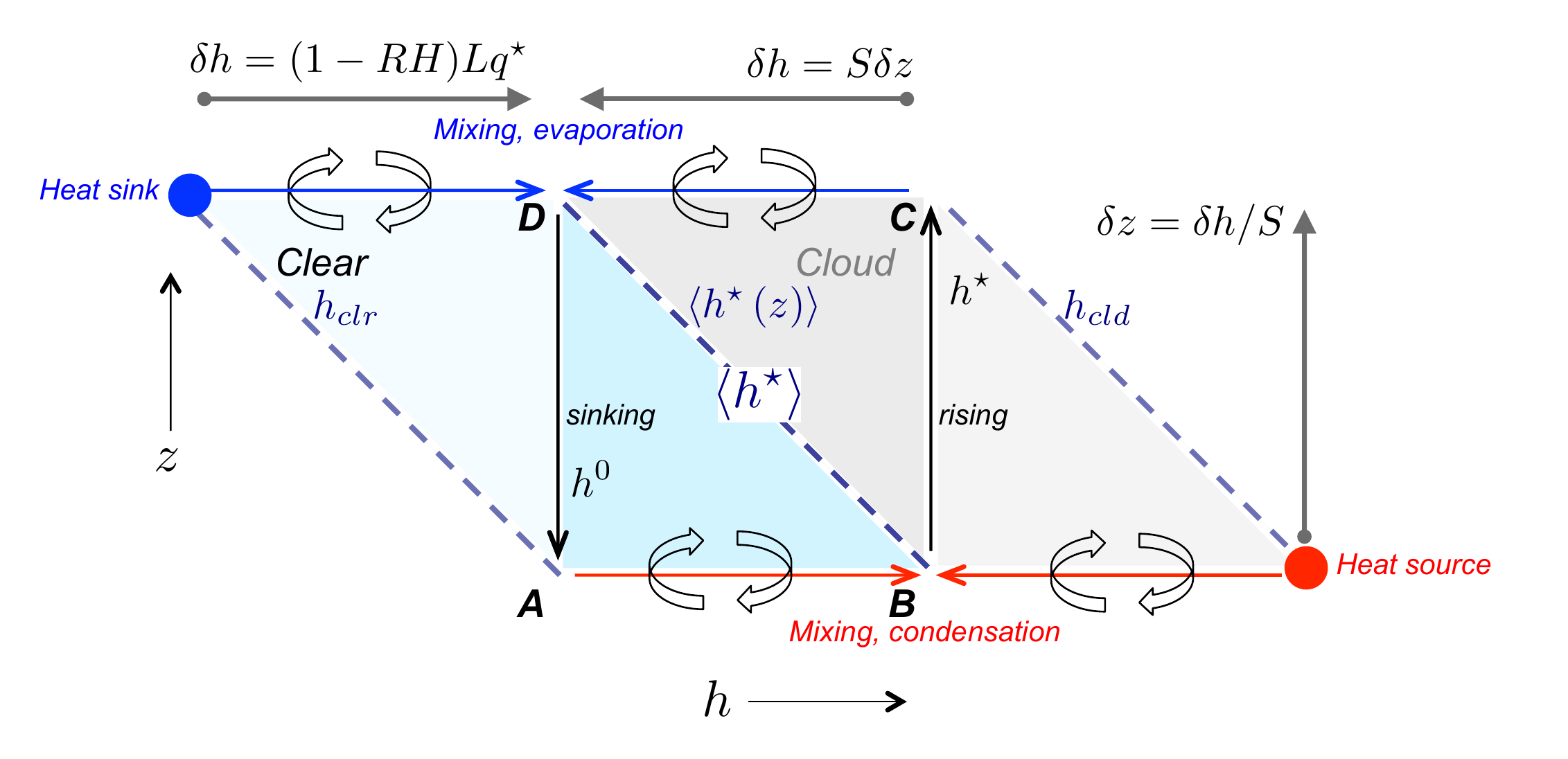}
\caption{\label{deltah} Illustration of a thermodynamic mixing engine along cloud boundaries in a convectively unstable atmosphere where $h^\star$ decreases with height. Mixing between the cloud and its environment at high (A to B) and low (C to D) values of constant $\left<h^\star\left(z\right)\right>$ leads to a circulation that is closed by cloudy moist adiabatic ascent (B to C) and clear-sky dry adiabatic descent (D to A). The energetic magnitude of the closed circulation is characterized by an equivalence between the potentials  associated with atmospheric stability $\delta h = S\delta z$ and relative humidity $\delta h = \left(1- RH\right)Lq^\star$. Note that the abscissa co-ordinate is in $h$ so the dashed blue lines $h\left(z\right)$ can be of any geometric shape. }
\end{figure}

A proposed thermodynamics for these processes is illustrated in the ``mixing engine'' diagram shown in Fig. \ref{deltah}. A somewhat similar description is the well-known Carnot cycle model of a hurricane  \citep{Emanuel1991}, where diabatic surface heat fluxes into the hurricane at constant $T$ are balanced by top-of-the-atmosphere thermal emission at a  colder temperature; the cycle is closed by moist adiabatic ascent at the hurricane core and dry adiabatic descent at its periphery. 

Here, however, circulations associated with entrainment and detrainment differ in that irreversible entropy production is due to material mixing rather than energetic exchange: the diabatic legs in the Carnot cycle that owe to radiative losses and heat gains at constant $T$ are replaced in the mixing engine by turbulent exchanges of air across cloud edge at constant $\left<h^\star\right>$. The contrasts in the atmosphere that sustain a reversible cycle are whatever large-scale atmospheric processes such as precipitation and top-of-the-atmosphere radiative cooling that maintain a steady-state lapse rate and relative humidity. 

\subsection{Horizontal exchanges} 
Fig. \ref{deltah} considers a convectively unstable moist atmosphere with $d\left<h^\star\right>/dz < 0$. To start, a sub-saturated air parcel outside cloud lies at a low altitude point A where $RH < 1$, where it has a temperature and height similar to the air lying at the edge of a nearby cloud located at point B. Air at A has a value of $h = h_{clr}$ and air at cloud edge is presumed to be neutrally buoyant with a saturated static energy of $\left<h^\star\right>$. It follows from Eq. \ref{eq:h} that the difference in moist static energy between the two air parcels at points A and B owes to their difference in relative humidity:
\begin{linenomath*}\begin{equation}
\delta h_{cld,clr} \equiv \left<h^\star\right> - h_{clr} = \left(1 - RH\right)L q^\star\left(h,z\right)\label{eq:deltah Delta RH}
\end{equation}\end{linenomath*}

Air lying on the cloud interior side of the cloud edge at the same level is saturated. It too has RH = 1, but it is more buoyant with a higher temperature. Accordingly, it has a moist static energy $h = h^\star$ that is greater than the value at cloud edge $\left<h^\star\right>$. If $h - \left<h^\star\right>$ and $\delta h_{cld,clr}$ are similar, then conservation of energy requires that any mixing by turbulence in equal parts of air at point A with this high $h$ air from the cloud interior will result in a just saturated mixed parcel at cloud edge with $h = \left<h^\star\right>$.  Through mixing, dry air becomes cloudy, so this diabatic event could be referred to as entrainment. Diabatic mixing of dry air with the saturated ``heat source" from the cloud interior likely displaces the cloud boundary in spatial co-ordinates, but with respect to $\left<h^\star\right>$, the cloud edge remains unchanged. 

\subsection{Vertical perturbations}
Following this mixing event, the cloud edge air parcel at point B is assumed to rise adiabatically along a moist isentrope within the cloud, maintaining a constant value of $h^\star$ as it moves upwards to point C. The average vertical profile in $\left<h^\star\right>$ is the atmospheric  stability given by:
\begin{linenomath*}\begin{equation}
S = \left|\frac{d\left<h^\star\right>}{d z}\right| \label{eq:Stability}
\end{equation}\end{linenomath*}
where $\left<h^\star\right>$ also serves as a neutral buoyancy point with respect to cloud edge. 

In Fig. \ref{deltah}, $d\left<h^\star\right>/dz < 0$, so air ascends within the cloud interior rather than along the cloud boundary. From Eq. \ref{eq:Stability}, the difference in the moist static energy between the air parcel and the cloud boundary is in proportion to the height above point B $\delta z$. Thus:
\begin{linenomath*}\begin{equation}
{\delta h_{stab}} \equiv h^\star - \left<h^\star\right> = {S}{\delta z} \label{eq:delta h Stability}
\end{equation}\end{linenomath*}
where $\delta z$ points in the direction of the circulation.

\subsection{A convective potential}
Motions of ascent in a cloud interior are turbulent. In the idealized cycle of Fig. \ref{deltah}, air at point C mixes with surrounding dry air to create a just-saturated mixture at point D located at cloud edge, thereby representing a detrainment event that is the counterpart to the earlier entrainment event between point A and B. The difference is that the mixing is with a sub-saturated ``heat sink" with low $h$ rather than a cloudy saturated ``heat source" with high $h$. Because the atmosphere is unstable, the value of $\left<h^\star\right>$ at point D is lower than that at point B. After this second mixing event, negatively buoyant clear air at point D that subsides a distance $\delta z$ along a dry adiabat to point A while maintaining constant $h^0$,
thereby completing the cycle of entrainment, moist ascent, detrainment, then dry descent. 

\begin{figure}
\centering
\includegraphics[width=0.95\linewidth]{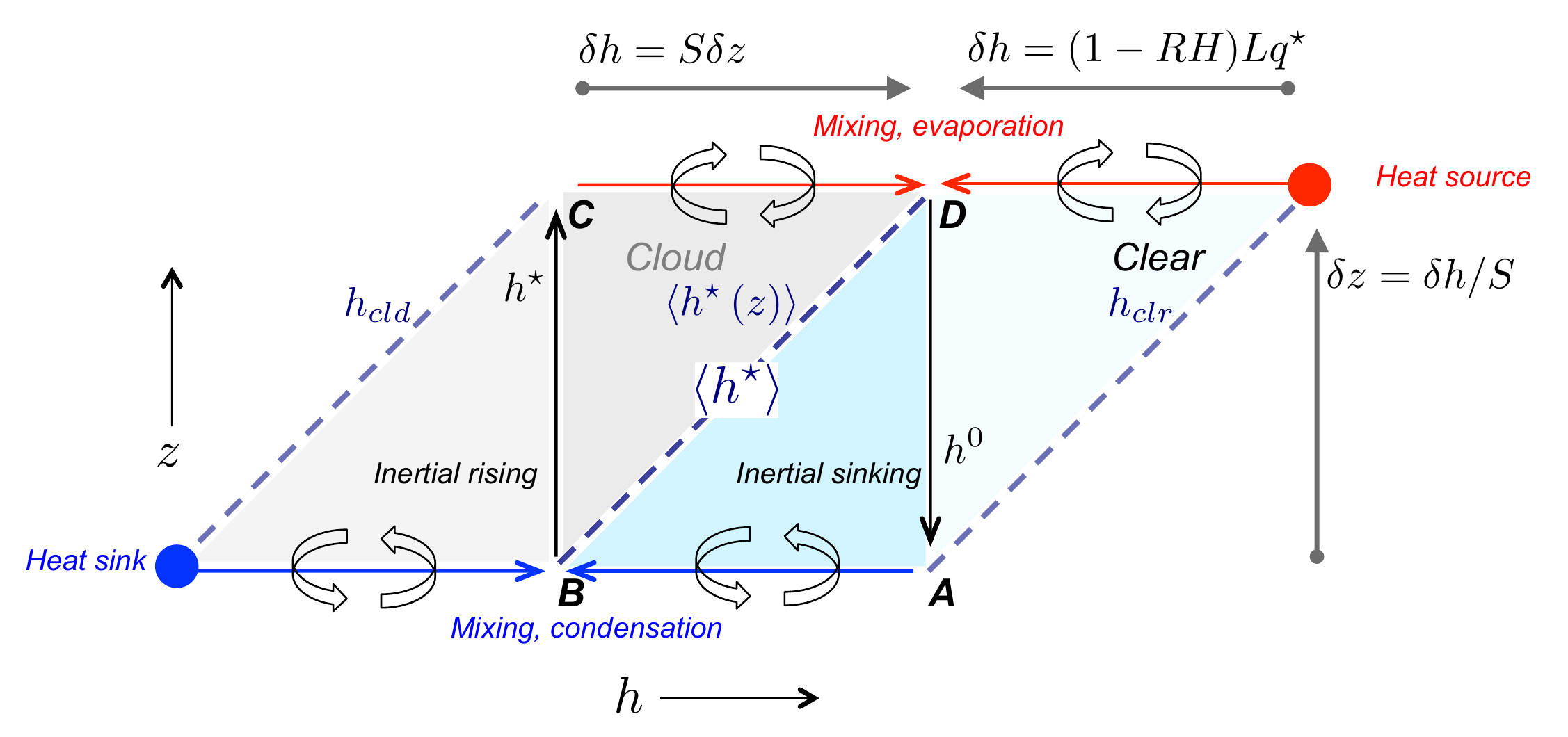}
\caption{\label{deltah stable} As for Fig. \ref{deltah} except the thermodynamic mixing engine lies within a convectively stable atmosphere where $h^\star$ increases with height. Diabatic mixing between the cloud and its environment at low (A to B) and high (C to D) values of constant $\left<h^\star\left(z\right)\right>$ does not spontaneously lead to a closed circulation as in the unstable case, except where vertical momentum overcomes stability to create the branches of moist adiabatic ascent (B to C) and, through continuity, clear-sky dry adiabatic descent (D to A). }
\end{figure}

Fig. \ref{deltah stable} shows the mixing engine within a stable rather than an unstable environment, where $d \left<h^\star\right>/dz > 0$. In this case, air at point B rises from a heat sink with low $\left<h^\star\right>$ to a heat source at C with higher $\left<h^\star\right>$. This process is not spontaneous as it requires an external source of potential energy to produce the vertical kinetic energy of ascent that overcomes static stability, originating perhaps in latent heat release lower down. The rising parcel mixes with its subsaturated environment to bring it into back into thermodynamic equilibrium with the cloud edge at the same level at point D. Here, the air has a higher value of $\left<h^\star\right>$ than at point B. Due to evaporative cooling, the air parcel then subsides outside the cloud to point A where it mixes with rising turbulent air in the cloudy interior to reform the cloud edge mixture at point B. 



Thus, whether the atmosphere is stable or unstable, and independent of cloud shape, the mixing engine formulation can be used to idealize buoyancy driven turbulent interactions at cloud edge in terms of the energetics of circulations along and across isentropic surfaces. Vertical isentropic displacement and lateral anisentropic mixing are part of a closed circulation of air with mass flux $J$ (units kg s$^{-1}$) about an equilibrium state $\left<h^\star\right>\pm\delta h/2$. Perturbations in $h$ related to turbulent moistening and drying (Eq. \ref{eq:deltah Delta RH}) are coupled to perturbations in $h$ due to cloud vertical motions (Eq. \ref{eq:delta h Stability}).  Since the circulation is closed, these two perturbations must be equal: at least on average, the interface between cloudy and clear skies in a cloud field can be defined by a generalized convective potential
\begin{linenomath*}\begin{equation}
\delta h = \delta h_{stab}= \delta h_{cld,clr} \label{eq:convective potential}
\end{equation}\end{linenomath*}
This equivalence represents the potential energy that is available to drive the circulations that turn clear air into cloudy air and vice versa. For cyclic motions around cloud edges, any potential energy that clouds gain during upward motion in an unstable atmosphere is ultimately lost through horizontal turbulent mixing with surrounding dry air. Dry descent completes the cycle.

It is interesting to consider the limiting case where the entire atmospheric profile follows a moist adiabat and $S=0$. Then, Eq. \ref{eq:convective potential} would imply the convective potential was zero, in which case the atmosphere would be everywhere cloudy. In fact, this special case has been employed elsewhere as the reference assumption for numerical convective adjustment schemes \citep{Arakawa1993}. The implication is that identifiable clouds, as defined by an interface between cloudy and clear air, can only exist if the atmosphere is on average sub-saturated with $\left<h\right> < \left<h^\star\right>$, whose temperature profile deviates from the moist adiabat.

\section{Timescales of circulations around cloud perimeter}

We now turn to the timescales of the circulations in the mixing engine, since these will be shown to be related to the length of the cloud perimeter. Implicit in the above discussion is an assumption that the cloud field lies in a quasi-equilibrium state; that is, the ``slow" timescales for changes in $\left<h^\star\left(z\right)\right>$ averaged over the atmospheric domain are much longer than the ``fast" timescales of local buoyancy driven perturbations about $\left<h^\star\left(z\right)\right>$. We are interested in examining a field of clouds over sufficiently large spatial scales and timescales that individual clouds and their lifecycles are resolved only as part of a larger ensemble, although not so long or large that $\left<h^\star\right>$ itself varies significantly as well \citep{ArakawaSchubert1974,CraigCohen2006}.
In general, this assumes that the external large-scale meteorological or radiative forcing that affects  the total mass, energy, and stability of the cloud field has characteristic scales of order days and hundreds of kilometers, depending on the synoptic meteorology \citep{Lord1980}. However, a prior estimate of the minimum domain required to satisfy this condition is of order 100 km across and an averaging time of approximately 1 hour \citep{Keane2012}. Here, we assume the fast timescale is related to the closed buoyancy circulations around cloud edges that form the mixing engine. The question is then, what is the ``fast" timescale, and how does it relate to cloud geometries?

 \subsection{Horizontal exchanges}

 Fick's Law for the diffusion of mass across an interface (units mass per time) is 
\begin{linenomath*}\begin{equation}
J = {\mathcal{D}\sigma}\nabla{\ln\rho h} \label{ref:diffusion_continuum}
\end{equation}\end{linenomath*}
where $\mathcal{D}$ is a diffusivity (units area per time) and $\sigma$ is the cross-section of the interface normal to the energy density gradient $\nabla{\rho h}$.  While Fick's Law is not specifically limited to molecular diffusion, a familiar example is the growth of a population $n$ of droplets of radius $r$ in response to net flows of water vapor molecules across the the interface between the droplet and its surroundings. For a small vapor density gradient, the net flow rate of water molecules normal to the total droplet surface is $J = 4\pi n r^2  \mathcal{D}\left<\rho\right>\nabla{e}/\left<e\right>$ where $e\propto \rho h$ is the vapor pressure (units energy density) and $\left<\rho\right>$ is the average vapor density. For a fixed vapor pressure difference $\Delta e$ in an infinite reservoir of vapor, the common expression is $J = 4\pi n r \mathcal{D}\left<\rho\right>\Delta{e}/\left<e\right>$  \citep{PruppacherKlett1997}.

Exchanges across cloud boundaries are driven primarily by turbulence, but the conceptual aspects of the mass transfer are the same \citep{Garrettmodes2012}: a cloud field can be treated as a population of particles that interacts with its clear-sky environment through a gradient in potential energy density across the cloud-edge interface; the rate of exchange is quantifiable through a diffusivity. Suppose a number distribution of cloud perimeters defined by the continuous functional form $n_{\lambda} \equiv dn/d\lambda$, so that the total perimeter of all clouds within a size range $\Delta \lambda$ with convective potential $\delta h$ is:
\begin{linenomath*}\begin{equation}
\Lambda = \int_\lambda^{\lambda + \Delta \lambda}n_\lambda\lambda d\lambda \label{eq:Lambda} 
\end{equation}\end{linenomath*}
Or, evaluated in discrete bins $j$, and applying the mean value theorem, the total perimeter per size bin  between $\lambda_j$ and $\lambda_j + \Delta\lambda_j$ with average size $\lambda_j$ is:
\begin{linenomath*}\begin{equation}
\Lambda_j = n_\lambda\left<\lambda\right>\Delta\lambda\simeq n_j\lambda_j\label{eq:Lambda MVT} 
\end{equation}\end{linenomath*}
where $n_j$ is the bin number. The total perimeter of all clouds in the cloud field ensemble is then $\Lambda = \sum_j n_j \lambda_j$. As a note, clouds tend to have fractal shapes \citep{lovejoy1982}, so the measured magnitude of $\lambda$ is a function of the ``ruler" length at which cloud edge is resolved $\xi$.

In terms of Fick's Law (Eq. \ref{ref:diffusion_continuum}), the physical interface between the ensemble of clouds and their surroundings is a disconnected strip with total area $\sigma = \Lambda\delta z = \Lambda \delta h/S$, where $S$ is the stability (Eq. \ref{eq:Stability}). Provided that $\delta h \ll \left<h^\star\right>$ (which is always the case in the atmosphere) and the atmospheric layer $\delta z$ has an average density $\left<\rho\right>$, and given that turbulent circulations around cloud boundaries are approximately isotropic \citep{Heus2008subsidingshells}, then $\nabla \ln \rho h \simeq \left<\rho\right> S$. Also, $\mathcal{D}$ is equal to the molecular diffusivity $\mathcal{D}_\eta$ only at scales smaller than the Kolmogorov microscale $\eta$ of $\sim$1 mm and due to turbulence. So, at coarser scales $\xi$ within the turbulent inertial subrange, an adjustment must be made:  
\begin{linenomath*}\begin{equation}
\mathcal{D} = \mathcal{D}_\eta\left(\xi/\eta\right)^{4/3}\label{eq:D_adjusted}
\end{equation}\end{linenomath*}
This upward modification to $\mathcal{D}_\eta$ compensates for the smaller resolved surface area of the fractal interface when it is measured at scales $\xi$ rather than $\eta$; the magnitude of energetic and material exchanges across a surface needs to be independent of the scale at which the surface is measured \citep{richardson1926,TennekesandLumley,Krueger1997}. 

Combined, these arguments (or, in fact, simple dimensional considerations) suggest a linearization of Eq. \ref{ref:diffusion_continuum} that leads to an expression for instantaneous exchanges across the interface between clouds and clear sky given by: 
\begin{linenomath*}\begin{equation}
J = {\mathcal{D}\left<\rho\right>}\Lambda \frac{\delta{h}}{\left<h^\star\right>} \label{eq:J_cloud}
\end{equation}\end{linenomath*}
where, as before, $\Lambda$ is related to the product of $n$ and $\lambda$. Eq. \ref{eq:J_cloud} can apply either to a single atmospheric layer or to the entire troposphere given that in either case the linearization condition $\delta h/\left<h^\star\right>\ll 1$ would be satisfied. With respect to horizontal exchanges, Eq. \ref{eq:J_cloud} can be expressed as  
\begin{linenomath*}\begin{equation}
J =  \frac{\left<\rho\right>V}{\tau}\frac{\left<\delta h^\star\right>}{\left<{h^\star}\right>} \label{eq:j differential}
\end{equation}\end{linenomath*}
where $V$ is the volume of the circulation and
\begin{linenomath*}\begin{equation}
\tau_{cld,clr}=V/\left(\mathcal{D}\Lambda\right)\label{eq:tau_Lambda}
\end{equation}\end{linenomath*} 
This timescale formulation is familiar elsewhere in the atmospheric sciences. Considering again the case of cloud droplets, suppose a total droplet concentration $n/V$ and average droplet radius $\left<r\right>$. The interfacial length density of the distribution determining the mass flux of condensation is $\Lambda/V = 4\pi\left<{r}\right>n/V$, and the phase relaxation time for dissipation of excess vapor is $\tau=V/\left(\mathcal{D}\Lambda\right)$ \citep{Kostinski2009}. The same phase relaxation time scale is seen in Eq. \ref{eq:tau_Lambda} for exchanges across cloud boundaries. That clouds are larger than droplets is not important for the energetics. Neither is the fact that cloud surfaces are rather more complex than droplets, since the same approach is applied to  dendritic ice crystals \citep{PruppacherKlett1997}. 

\subsection{Vertical perturbations}
 From the discussion in Sec. \ref{sec:MSE differences}, prior theoretical and modeling work suggests that buoyant accelerations $b$ on either side of the edge of a cloud are closely related to local gradients in moist static energy, in which case, to a good approximation:
\begin{linenomath*}\begin{equation}
b  \simeq g\delta h / \left<h^\star\right> \label{eq:b}
\end{equation}\end{linenomath*}
In terms of an adiabatic displacement $\delta z$ in the direction of the circulations in a mixing engine, as in from points B to C and D to A in Fig. \ref{deltah}, these buoyant accelerations around cloud edge can be expressed as:
\begin{linenomath*}\begin{equation}
b = \frac{d^2 \delta{z}}{d t^2} = {N^{\star}}^2 \delta{z} \label{eq:buoyancy oscillations}
\end{equation}\end{linenomath*}
where ${N^\star}^2\delta z$ is the restoring force (per unit mass) and $N^\star$ is a frequency. If $\delta h = S\delta z$ with respect to $\left<h^\star\right>$ at cloud edge (Eq. \ref{eq:delta h Stability}), it follows from equating Eqs. \ref{eq:b} and \ref{eq:buoyancy oscillations} that:  
\begin{linenomath*}\begin{equation}
N{^\star}^2 = \frac{Sg}{\left<h^\star\right>} \label{eq:buoyancy oscillations N}
\end{equation}\end{linenomath*}
 
 Eq. \ref{eq:buoyancy oscillations N} provides a timescale for the rising and falling motions of air parcels in the mixing engine equal to 
\begin{linenomath*}\begin{equation}
\tau_{stab} \sim \frac{1}{N^\star} = \sqrt{\frac{\left<h^\star\right>}{Sg}}\label{eq:tau_N}
\end{equation}\end{linenomath*}
A related quantity to $N^\star$ is the Brunt-V\"{a}is\"{a}l\"{a} frequency for dry adiabatic oscillations in a stable environment defined by $N^2 = \left(gd\left<h^0\right>/dz\right)/\left(c_p T\right)$. For the tropical troposphere taken as a whole,  $N \sim 0.01$\,s$^{-1}$ \citep{mapes2001water}. While $N$ is a different expression than that provided for $N^\star$ (Eq. \ref{eq:buoyancy oscillations N}), an estimate for $\tau$ of order $1/N = 100$ seconds should provide a rough guide for how fast mixing occurs across cloud boundaries within the larger cloud field: as shown in Fig. \ref{Isosurface}, $d\left<h\right>/dz$ is only slightly steeper than $d\left<h^\star\right>/dz$.

\subsection{The relationship of cloud perimeter to stability}



Expressing the instantaneous magnitude of circulations $J$ around cloud edge using an orthogonal reference frame of number $n$, perimeter $\lambda$, and convective potential $\delta h$ has the specific advantage that the physics is agnostic to the complexities of the cloud spatial structure. Additionally, as exemplified in Fig. \ref{MSE_ISO_and_PDF}, there is no need to consider the complication that any given $h$ surface can be found at two different heights in a convective atmosphere. $J$ applies to circulations within layers around a value of $\left<h^\star\right>$ independent of vertical location. There is no requirement of modeling the spatial advection of air between two vertical levels since these are the same isentrope in $h^\star$. 

This abstraction also leads to a useful link between tropospheric bulk thermodynamics and cloud morphology. We have previously argued that cloud edge air circulations in an idealized mixing engine involve potential energy contrasts due to horizontal latent heat and vertical stability gradients that are equivalent, such that $\delta h_{stab} = \delta h_{cld,clr}$ is the convective potential $\delta h$ (Eq. \ref{eq:convective potential}).  Within the same framework, we can assume that the rate of potential energy production and dissipation in local turbulent circulations $\delta h/\tau$ is also independent of lateral and vertical direction, in which case there is an expected equality between the timescale for vertical buoyancy oscillations (Eq. \ref{eq:tau_N}) and the phase relaxation timescale for horizontal exchanges across cloud boundaries (Eq. \ref{eq:tau_Lambda}). That is:
\begin{linenomath*}\begin{equation}
\tau = \tau_{stab} =  \tau_{cld,clr} \label{eq:timescale equivalence}
\end{equation}\end{linenomath*}
This leads to the result that the total cloud perimeter and the buoyancy frequency are related by: 
\begin{linenomath*}\begin{equation}
\Lambda = \frac{V}{\mathcal{D}}N^\star \label{eq:Dlambda_N}
\end{equation}\end{linenomath*}
or, with respect to the stability 
\begin{linenomath*}\begin{equation}
\Lambda = \frac{V}{\mathcal{D}}\sqrt{\frac{Sg}{\left<h^\star\right>}}\label{eq:Dlambda_S}
\end{equation}\end{linenomath*}
The total perimeter of a cloud field of volume $V$, resolved horizontally and vertically at spatial scale $\xi$, scales as the square root of the tropospheric stability.



\section{Statistics of cloud perimeters}

Eq. \ref{eq:Dlambda_S} links the total perimeter of clouds in a cloud field $\Lambda$ to the tropospheric stability with respect to a moist adiabat  $S$. The next question is how the total perimeter within a cloud field volume is distributed statistically among individual clouds. We consider two cases, the first being the number distribution of cloud perimeters for a fixed convective potential $\delta h$, and the second the number distribution as a function of convective potential.  Effectively, the discussion shifts from the more usual consideration of the temporal evolution of cloud areas as a function of height to the purely statistical consideration of cloud geometries in a revised co-ordinate system of $\delta h$, $n$, and $\lambda$.

\subsection{Perimeter distributions for a fixed convective potential}

\begin{figure}
\centering
\includegraphics[width=.95\linewidth]{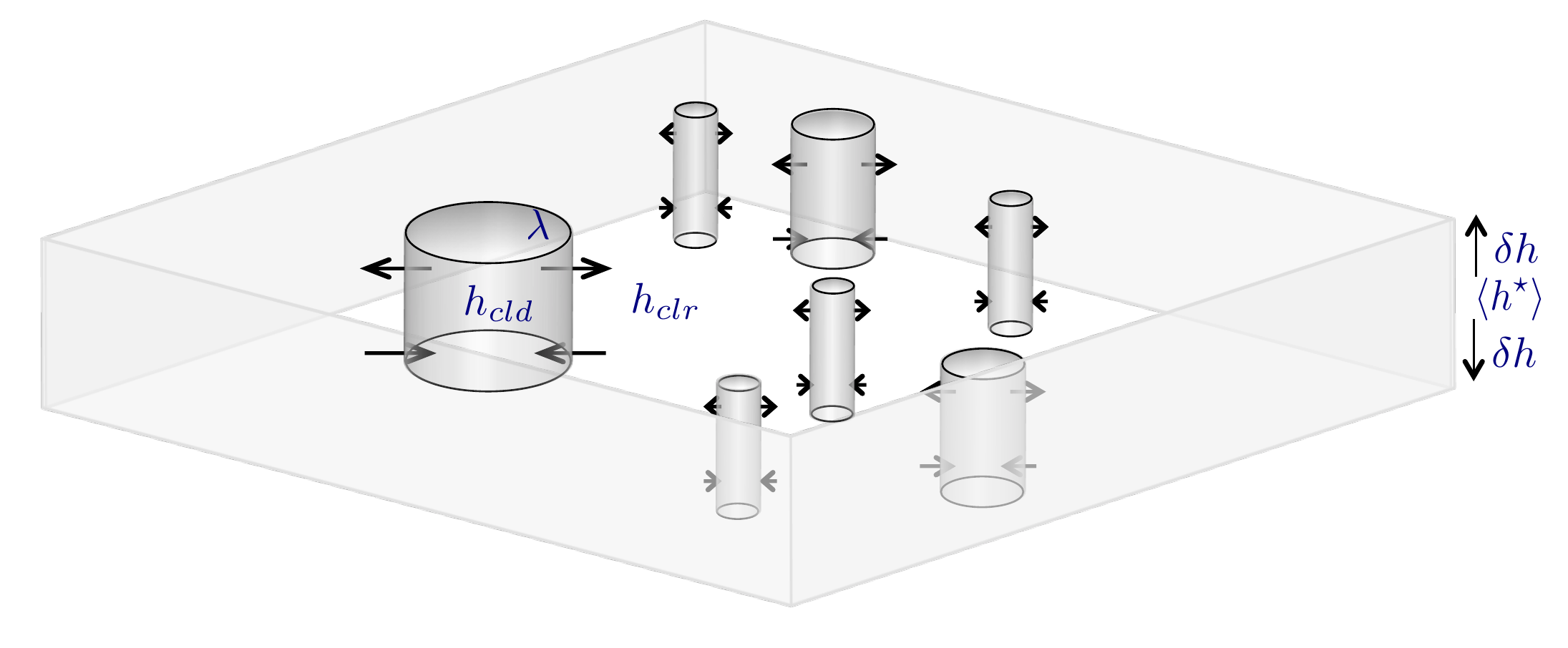}
\caption{\label{fig:Microcanon}Idealization of cloud field circulations within a moist isentropic layer with mean saturated static energy $\left<h^\star\right>$ at the interface between clouds and clear-skies defined by a range of cloud perimeters $\lambda$}
\end{figure}

\begin{figure}
\includegraphics[width=.95\linewidth]{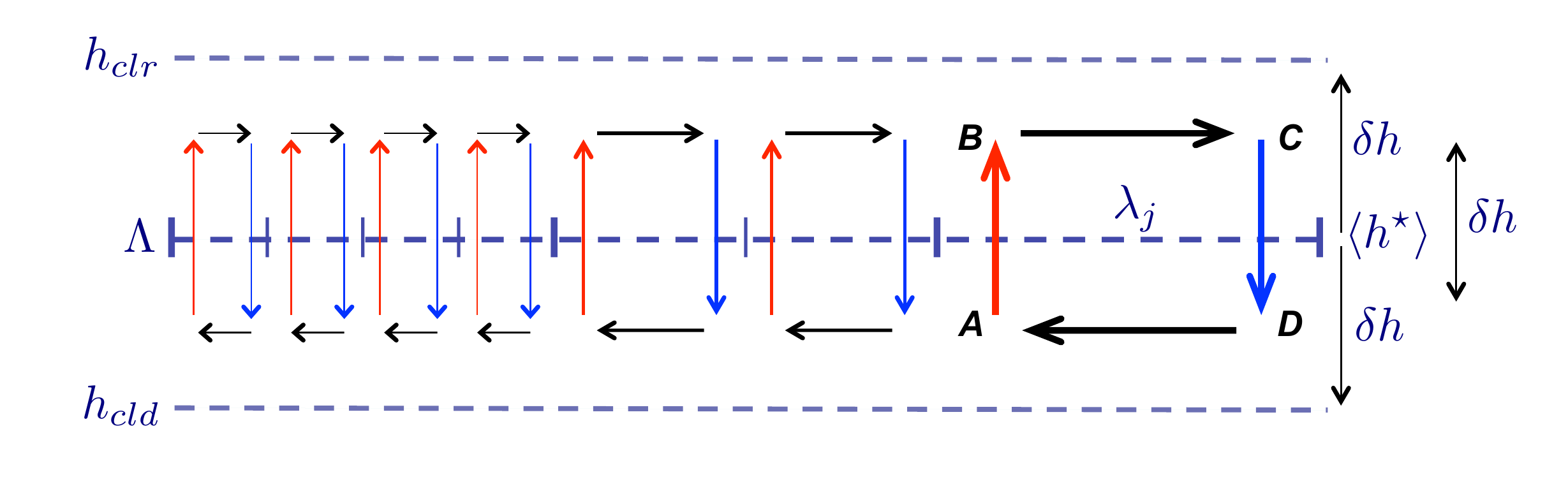}
\caption{\label{fig:AlongIsentropes}An alternative representation of Fig. \ref{fig:Microcanon}, with distributions of cloud perimeter $\lambda_j$ within an atmospheric isentropic layer of fixed $\delta h$ and associated circulations $J_j$ about $\left<h^\star\right>\pm\delta h/2$. Cloud distributions are divided to satisfy the constraint that the total perimeter is given by $\Lambda = \sum_j n_j \lambda_j$. The schematic illustrates the power law relationship that size classes that are twice as small are twice as common with the same total circulation. Letters correspond to points in the mixing engine illustrated in Fig. \ref{deltah} }
\end{figure}



%

We start with Eq. \ref{eq:j differential} by defining a system where all air along cloud edge has the same convective potential $\delta h$ and saturated static energy $\left<h^\star\right>$, or that we are considering an isentropic layer of physical thickness $\delta z = \delta h/S$ with uniform density $\left<\rho\right>$. We assume that the entire volume of air $V$ has had sufficient time to undergo all possible exchanges between clear skies and clouds, i.e  the ergodic condition is satisfied whereby no set of circulations around cloud perimeters has any privileged possession to any particular volume or circulation of air over the long-term.



From Eqs. \ref{eq:j differential} and \ref{eq:tau_N}, the total cloud edge circulation rate $J$ is determined by the atmospheric static stability, which in turn can be linked to the total cloud perimeter. Thus, $\Lambda$ is a conserved state variable of the ensemble that relates $\delta h$ to $J$.  Figures \ref{fig:Microcanon} and \ref{fig:AlongIsentropes} show idealized representations of how $\Lambda$ and $J$ might be partitioned among clouds of varying perimeter classes $\lambda_j$. What is assumed is a fixed range of values of $h = \left<{h}^\star\right>\pm\delta h$ with clear skies defined  by $h_{clr}<\left<h^\star\right>$, cloudy edges by $\left<h^\star\right>$, and cloud interiors by $h_{cld}>\left<h^\star\right>$. The convective potential is $\delta h = \left<h^\star\right> - h_{clr}$. From Eqs. \ref{eq:Lambda MVT} and \ref{eq:J_cloud}, cloud edge circulations are partitioned among size classes according to:
\begin{linenomath*}\begin{equation}
J_j = {\mathcal{D}\left<\rho\right>}\frac{\delta{h}}{\left<h^\star\right>}n_j\lambda_j  \label{eq:J_cloud_j}
\end{equation}\end{linenomath*}

Individual clouds in a cloud field constantly change their shape through condensation, entrainment, and merging. However, statistically speaking, there are global constraints on what is possible. Suppose, by way of example, a growth event that translates a number of clouds by whatever mechanism from coordinates $\left[n_{j-1}, \lambda_{j-1}, \delta h \right]$ to $\left[n_{j}, \lambda_{j}, \delta h \right]$. To satisfy steady-state in stability and therefore $J$, from Eq. \ref{eq:J_cloud}, the total perimeter $\Lambda = \sum_j n_j \lambda_j$ would need to be conserved. This would require that the product of number and perimeter $n_j\lambda_j$ would increase by the same amount in size bin $j$ that it decreased in size bin $j-1$. From Eq. \ref{eq:J_cloud_j}, the volume of circulations around cloud edge would shift to size class $\lambda_j$ at the expense of those in size class $\lambda_{j-1}$ .  

Stationarity in the size distribution could be maintained if there were an irreversible cascade of circulating air from one size class to the next at an equal rate $J$ (units mass per time), analogous to the manner in which turbulent kinetic energy is passed progressively to ever smaller eddies in the inertial subrange at constant rate $\varepsilon$ (units energy per time). For there to be no net flux convergence of air within any given size bin $\lambda_j$, this would require that $dJ/d\lambda = 0$. From the differential form of Eq. \ref{eq:J_cloud_j}:
\begin{linenomath*}\begin{equation}
\frac{dJ}{d\lambda} = {\mathcal{D}\left<\rho\right>}\frac{\delta{h}}{\left<h^\star\right>}\frac{d\left(n\lambda\right)}{d\lambda} = 0
\end{equation}\end{linenomath*}
This leads to the functional form for the size distribution: 
\begin{linenomath*}\begin{equation}
n_\lambda = -n/\lambda\label{eq:size distribution}
\end{equation}\end{linenomath*}
or in logarithmic space
\begin{linenomath*}\begin{equation}
\frac{d\ln n}{d\ln{\lambda}}=-1\label{eq:power law}
\end{equation}\end{linenomath*}
For $\nu$ equally logarithmically spaced bins with average size $\lambda_j$, $n_j\lambda_j$ is a constant independent of $\lambda_j$ whereby 
\begin{linenomath*}\begin{equation}
n_j\lambda_j = \Lambda/\nu \label{eq:constant bins}
\end{equation}\end{linenomath*}
Thus, for constant $\delta h$, as shown in Figures \ref{fig:Microcanon} and \ref{fig:AlongIsentropes}, clouds have the scale invariant property for steady-state size distributions given by the power-law relating number and perimeter $n_j\propto\lambda_j^{-1}$.

An important nuance to the cascade argument is that, where with turbulent energy dissipation eddies at one size extreme get converted through viscous forces to heat, the aforementioned cloud cascade is not unidirectional; once formed, the largest clouds (e.g., cirrus anvils) do not simply disappear into some non-cloudy state. Instead, stationarity requires a simultaneous reverse cascade whereby clouds evaporate, detrain, and split. Naturally, these processes are highly interactive and complicated. Fortunately, there is no need to appeal to any of the specifics of the physics to obtain the size distribution other than the steady-state constraint that all cloud size classes compete for the same flows of air, and therefore the same total perimeter.

\subsection{Perimeter distributions as  a function of  convective potential}\label{sec: across isentropes}
\begin{figure}
\includegraphics[width=.95\linewidth]{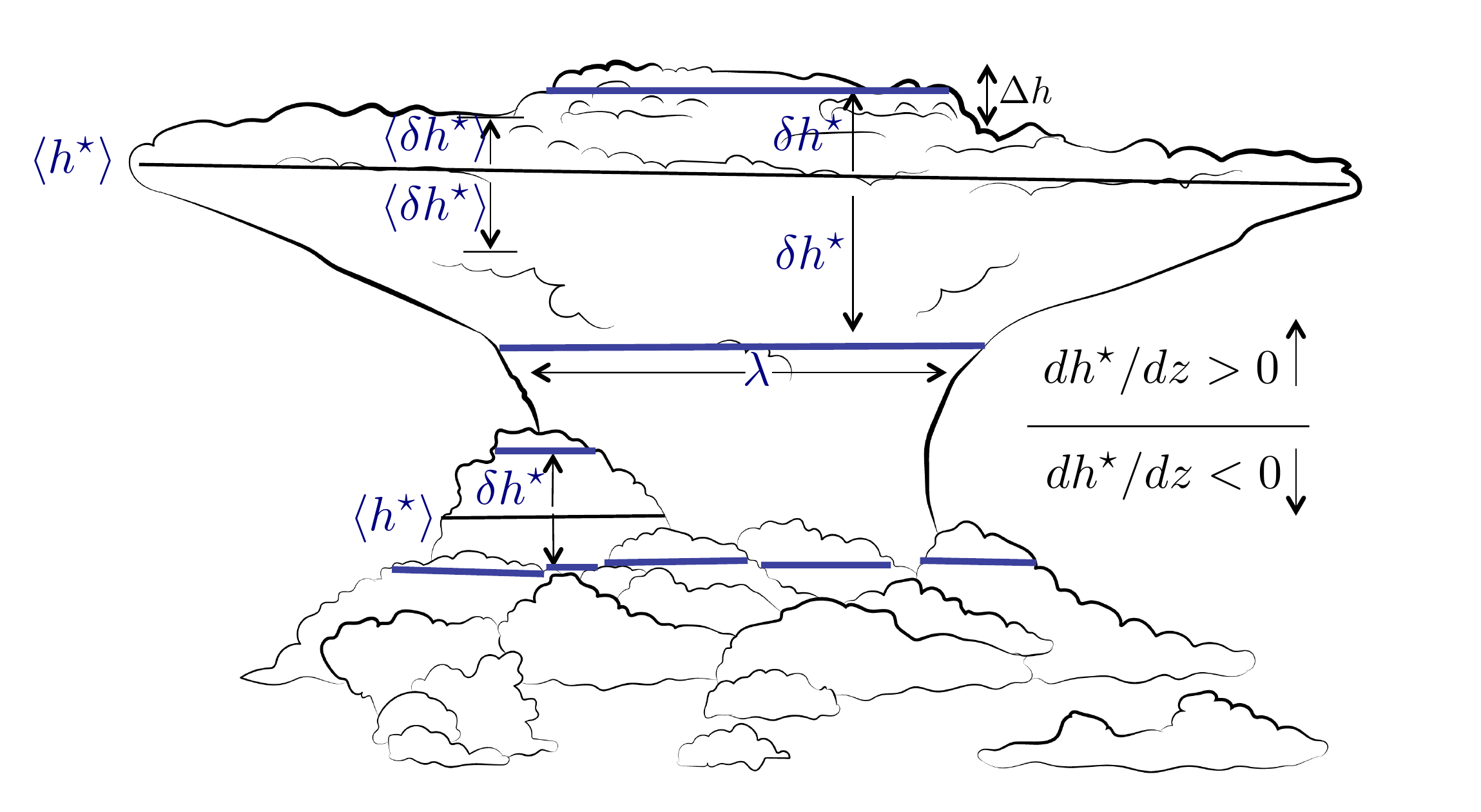}
\caption{\label{fig:Cbcartoon}Cartoon of perturbations $\delta h^\star$ from an ensemble mean $\left<h^\star\right>$ of a field of clouds with atmospheric stability $dh^\star/dz$, considered within a prespecified range of values of saturated static energy $\Delta h$. $\left<\delta h^\star\right>$ is the standard deviation of $\delta h^\star$  for all cloud edges. $\lambda$ is the perimeter of a cloud at any given value of $h^\star$.}
\end{figure}

Having considered how cloud geometry and air circulations vary statistically within an atmosphere defined by a fixed convective potential we now consider how total perimeter varies with  convective potential. 
As illustrated in Fig. \ref{fig:Cbcartoon},  atmospheric fluctuations that enable the formation of an interface separating clouds and clear air provide the energy that allows air along cloud edges to deviate from its neutrally-buoyant equilibrium state $h = \left<h^\star\right>$ by an amount equal to the convective potential $\delta h = S\delta z$. Because air at cloud edges is always saturated, this disequilibrium is expressible in terms of the saturated static energy as $\delta h^\star = S \delta z = h^\star - \left<h^\star\right>$. 

The disequilibrium $\delta h^\star$ leads to associated mixing between clouds and surrounding clear air, as in Fig. \ref{deltah}. The total of all cloud edge circulations, obtained by integrating the total perimeter distribution $\Lambda_{\delta h^\star} = d\Lambda/d\delta h^\star$ over all associated values of convective potential $\delta h^\star$ (Eq. \ref{eq:J_cloud}) is: 
\begin{linenomath*}\begin{equation}
J = \frac{\mathcal{D}\left<\rho\right>}{\left<h^\star\right>}\int_0^\infty\Lambda_{\delta h^\star}\delta h^\star d\delta h^\star \label{eq:J_cloud_integral}
\end{equation}\end{linenomath*}
or, evaluated in discrete bins $i$, as for Eq. \ref{eq:J_cloud_j}:
\begin{linenomath*}\begin{equation}
J_i = \frac{\mathcal{D}\left<\rho\right>}{\left<h^\star\right>}\Lambda_{\delta h^\star}\delta h^\star_i\Delta h\simeq \frac{\mathcal{D}\left<\rho\right>}{\left<h^\star\right>} \Lambda_i\delta h_i^\star\label{eq:J MVT} 
\end{equation}\end{linenomath*}
where $\Lambda_i$ is the total perimeter of clouds with average departure from equilibrium $\delta h_i^\star$ calculated in the interval $\delta h_i^\star \pm \Delta h/2$ at horizontal and vertical resolution $\xi$:
\begin{linenomath*}\begin{equation}
\Lambda_i = \sum_{j} n_{i,j} \lambda_{j} \label{eq:Lambda_sum}
\end{equation}\end{linenomath*}
For the situation that the convective potential $\delta h$ was prescribed, a power-law solution was obtained for $n_\lambda$ by assuming that stationarity in the size distribution requires zero convergence of air in size classes $\lambda$ such that $dJ/d\lambda = 0$. We now take the same approach with respect to varying $\delta h^\star$ to derive $\Lambda_{\delta h^\star}$ by assuming that $dJ/d{\delta h^\star} = 0$.

As previously, the starting point is to allow that there is sufficient time compared to convective mixing times $\tau$ that the entire atmospheric mass $m$ and volume $V$ cycles through all available states in $\Lambda$ and $\delta h^\star$, i.e. the ergodic condition. Then, from Eq. \ref{eq:J MVT}, any individual event that increases the convective potential of a given circulating air mass at cloud edge must eventually take away from the potential energy of air that is available for circulations elsewhere. Stationarity in the total perimeter distribution is only possible provided that there is a continuous throughput of air $J$ through $\delta h^\star$ so that there is no net flux convergence at any given potential $\delta h^\star$. Thus: 
\begin{linenomath*}\begin{equation}
\frac{dJ}{d\delta h^\star} = \frac{\mathcal{D}\left<\rho\right>}{\left<h^\star\right>}\left(\delta h^\star\frac{d\Lambda}{d\delta h^\star} + \Lambda\right) = 0 \label{eq: J heating working}
\end{equation}\end{linenomath*}
in which case, it follows that 
\begin{linenomath*}\begin{equation}
\Lambda_{\delta h^\star}  = - \frac{\Lambda}{\delta h^\star} \label{eq:lambda dh classes steady state}
\end{equation}\end{linenomath*}

Eq. \ref{eq:lambda dh classes steady state} would appear to lead to a power law as in Eq. \ref{eq:power law}, except that it is clear from Figs. \ref{MSE_ISO_and_PDF} and \ref{Isosurface} that atmospheric variability in $\delta h^\star$ is much less than variability in cloud perimeter. As a measure of the average magnitude of departures from the mean, the standard deviation is commonly used. More specifically, along cloud edges, variability can be expressed as:
\begin{linenomath*}\begin{equation}
\left<\delta h^\star\right> = \sqrt{\left<\left(h^\star - \left<{h^\star}\right>\right)^2\right>} \label{eq:square root}
\end{equation}\end{linenomath*}
If we assume that $\left<\delta h^\star\right>$ is quasi-independent of $\delta h^\star$, then, from  Eq. \ref{eq:lambda dh classes steady state}
\begin{linenomath*}\begin{equation}
\frac{d\ln\Lambda}{d\delta h^\star} \simeq -\frac{1}{{\left<\delta h^\star\right>}} \label{eq:dlnlambda} 
\end{equation}\end{linenomath*}
Eq. \ref{eq:dlnlambda} implies that the relative change in perimeter from one isentrope to another is constant, and proportional to the inverse of the standard deviation from the mean given by Eq. \ref{eq:square root}. Integrating Eq. \ref{eq:dlnlambda} between $\left[\Lambda_0,0\right]$ and $\left[\Lambda, \delta h^\star\right]$ we obtain a negative exponential or Boltzmann distribution:
\begin{linenomath*}\begin{equation}
{\Lambda} = \Lambda_{tot}\exp\left(-\beta{\delta h^\star}\right) \label{eq:lambda delta h} 
\end{equation}\end{linenomath*}
where
\begin{linenomath*}\begin{equation}
\beta = \frac{1}{\left<\delta h^\star\right>} \label{eq:beta} 
\end{equation}\end{linenomath*}
is the inverse temperature of statistical mechanics and
\begin{linenomath*}\begin{equation}
\Lambda_{tot} = \int_0^\infty \frac{d\Lambda}{d\delta h^\star} d\delta h^\star \label{eq:Lambda tot}
\end{equation}\end{linenomath*}
Expressed in discrete energy bins of width $\Delta h$, where $\delta h_i^\star = \left(i - 1/2\right) \Delta h$ for $i \geq 1$, the total perimeter at $\delta h^\star $ is:  
\begin{linenomath*}\begin{equation}
\Lambda_i  =  \int^{\delta h_i^\star + \Delta h/2}_{\delta h_i^\star - \Delta h/2}\frac{d\Lambda}{d\delta h^\star}d\delta h^\star \label{eq:Lambda i integral} 
\end{equation}\end{linenomath*}
so that the integral expression Eq. \ref{eq:Lambda tot} can be re-expressed in discretized form as:
\begin{linenomath*}\begin{equation}
\Lambda_{tot} = \sum_i\Lambda_i \label{eq:Lambda tot discrete}
\end{equation}\end{linenomath*}
Then, from Eqs. \ref{eq:lambda delta h} and \ref{eq:Lambda tot}, 
\begin{linenomath*}\begin{equation}
\Lambda_i   = \Lambda_{0}\exp\left(-\beta{\delta h_i^\star}\right)\label{eq:Lambda_i} 
\end{equation}\end{linenomath*}
where
\begin{linenomath*}\begin{equation}
\Lambda_0    = \Lambda_{tot}\left(1 - \exp\left(-\beta{\Delta h}\right)\right) \label{eq:Lambda_0} 
\end{equation}\end{linenomath*} 
So, for a case where discrete data for $\Lambda_i$ versus $\delta h_i^\star$ are plotted on a log-linear plot, the expected slope of the line would be $-\beta = -1/\left<\delta h^\star\right>$ and the intercept at  $\delta h_i^\star = 0$ equal to $\Lambda_0$. In the limit that $\Delta h/\left<\delta h^\star\right>\ll1$, the intercept could be simplified to  $\Lambda_0 = \Lambda_{tot}\beta{\Delta h_i}$.  

Because $\Lambda_i$ follows a Boltzmann distribution with respect to $\delta h^\star_i$ (Eq. \ref{eq:Lambda_i}), it follows from Eq. \ref{eq:Lambda_sum} that a Boltzmann distribution should also be expected to apply to the populations of clouds $n_{i,j}$ in any fixed size bin $\lambda_{j}$. Thus:
\begin{linenomath*}\begin{equation}
n_{i,j}    = n_{0,j}\exp\left(-\beta{\delta h_i^\star}\right)\label{eq:lambda_i} 
\end{equation}\end{linenomath*}
where
\begin{linenomath*}\begin{equation}
n_{0,j}    = n_{tot,j}\left(1 - \exp\left(-\beta{\Delta h}\right)\right) \label{eq:n_0} 
\end{equation}\end{linenomath*} 

\subsection{Perimeter distributions derived from bulk thermodynamic stability}

 Eq. \ref{eq:Dlambda_S} suggests that $\Lambda_{tot}$ can be inferred from a measure of the tropospheric stability $S$. Additionally, $\Lambda_{tot}$ is the sum of the total cloud perimeter summed over bins of convective potential (Eq. \ref{eq:Lambda tot discrete}), where the total perimeter follows a negative exponential with respect to the departure of the convective potential $\delta h^\star$ from the equilibrium value for the entire cloud field domain $\left<h^\star\right>$ (Eq. \ref{eq:Lambda_i}). Within each of these bins, dividing the total perimeter into $\nu$ logarithmically spaced bins in individual cloud perimeter, the size distribution obeys a power-law given by Eq. \ref{eq:power law}. Thus, from Eqs. \ref{eq:Dlambda_S}, \ref{eq:constant bins}  and \ref{eq:Lambda tot discrete} to \ref{eq:n_0}, the number distribution of clouds is expected to follow: 
 \begin{linenomath*}\begin{equation}
n_{i,j} = \frac{V}{\mathcal{D}}\sqrt{\frac{Sg}{\left<h^\star\right>}}\frac{\left(1 - \exp\left(-\beta{\Delta h}\right)\right)}{\nu}\frac{\exp\left(-\beta{\delta h_i^\star}\right)}{\lambda_j}\label{eq:gamma distribution}
 \end{equation}\end{linenomath*} 
 Thus, perhaps rather remarkably, it appears the average bulk stability of the atmospheric volume may contain sufficient information with which to obtain statistical distributions for cloud geometries along or across moist isentropic surfaces, provided a suitable estimate can be provided for $\beta = 1/\left<\delta h^\star\right>$.

\section{Comparison with numerical simulations}


The relationships for cloud forms we have derived are now compared with the much more spatially and temporally complex output from the Giga-LES, which is assumed to serve as a simulated ``truth''. We obtain the perimeter along moist isentropic surfaces in the simulation by first identifying as cloudy any grid point that has a non-precipitating condensate mixing ratio in excess of 101\% of $q^\star\left(T, p\right)$ \citep{krueger-1995-improvements}. We then identify contiguous 2D groups of cloudy grid points along horizontal slices at each vertical level. Points that define the edges of each group are traced to find the perimeter length for each cloud.  

The perimeters of holes in clouds are considered as well as those of clouds in a clear-sky environment. For computational reasons related to the grid structure of the model, we evaluate cloud perimeter values along height surfaces  rather than surfaces of $h^\star$; less than 4\% of cloud perimeters calculated at a given altitude deviate from the mean value of $h^\star$ by more than 1\,kJ\,kg$^{-1}$, and these perimeters were discarded from the analysis. The saturated static energy $h^\star$ at each contiguous point along the perimeter at that altitude is then averaged to yield $\lambda\left(h^\star\right)$ for the cloud. The proxy calculations implicitly account for the fact shown in Figure \ref{MSE_ISO_and_PDF} that clouds have isentropic surfaces that can be multi-valued with height, i.e., that $\Lambda$ can be single-valued in $h^\star$ but bi-valued in $z$. Further, perimeters are calculated at a fixed horizontal and vertical resolution of $\xi = 100$\,m, proportionately sub-sampling at lower altitudes where vertical resolutions are higher. Analysis focuses on 12 hours of simulation after 12 hours of model spin-up time in the full Giga-LES simulation.

\begin{figure}
\centering
\includegraphics[width=1.1\linewidth]{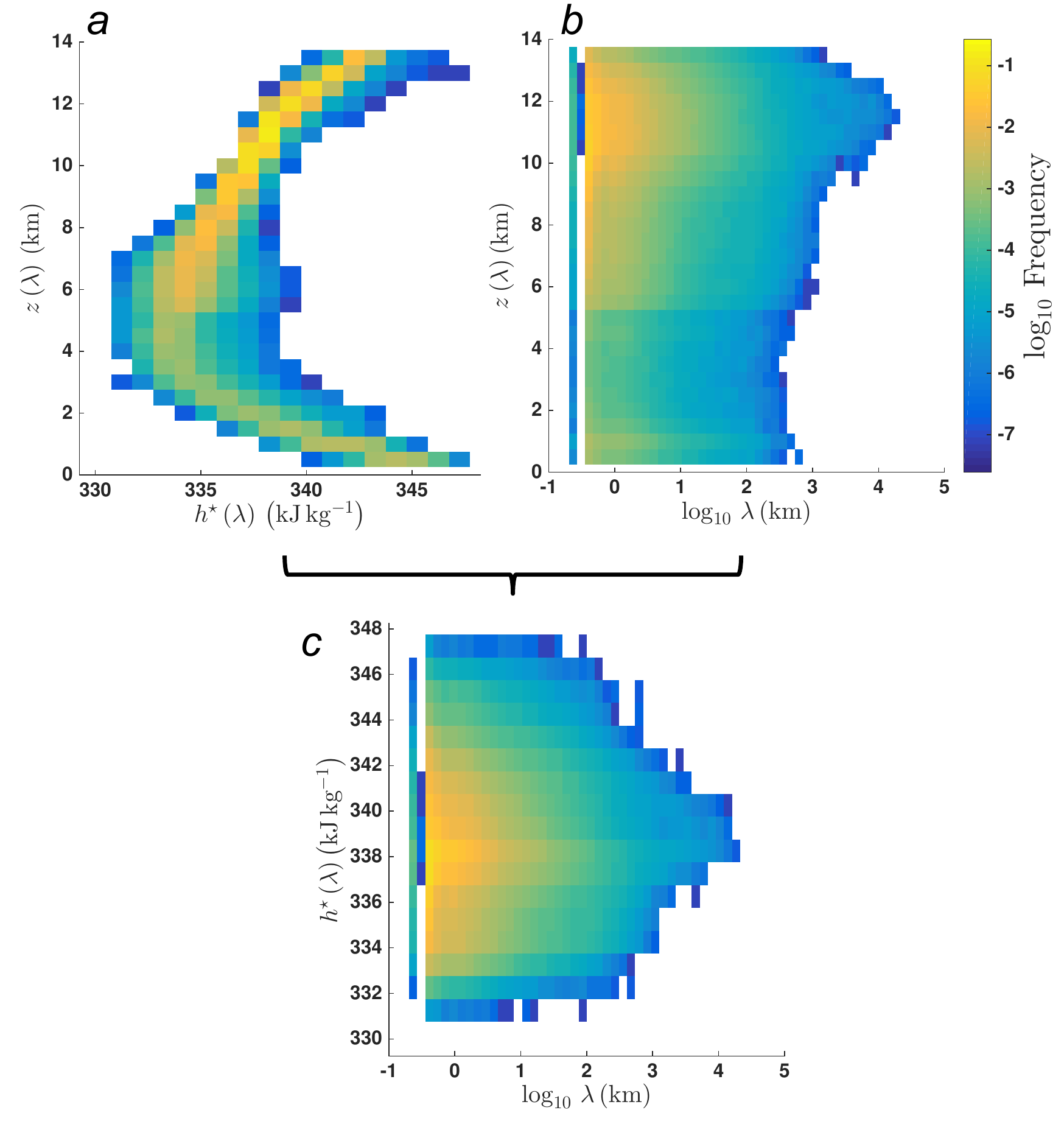}
\caption{\label{fig:3distributions}Giga-LES normalized frequency distributions evaluated along cloud perimeters of (a) saturated static energy versus height (b), perimeter versus height and (c) saturated static energy versus perimeter. }
\end{figure}

Frequency distributions with respect to $\lambda$, $h^\star$ and $z$ are shown in Fig. \ref{fig:3distributions}. In total, 99\% of cloud perimeters lie below 13.2\,km height with a mean value and standard deviation $\left<h^\star\right> \pm \left<\delta h^\star\right>$ of  $337.5 \pm 1.3$\,kJ\,kg$^{-1}$. With respect to height, $h^\star$ has a parabolic form, with instability with respect to a moist adiabat below approximately 5\,km altitude and stability above. Almost all cloud edge cells lie within the stable portion, representing 97\% of the total, with a mean altitude of 10.5$\pm$1.5\,km. The largest perimeter clouds with $\lambda > 1000$\,km are only found in the stable portion, with an average altitude of 11.2$\pm$0.8\,km.

Fig. \ref{fig:3distributions}c illustrates the utility of representing the full spatial complexity of cloud shapes within this revised coordinate system of $h^\star$ and $\ln\lambda$ since it reduces to frequency distributions that are nearly symmetric about the mean value $\left<h^\star\right>$. Frequencies drop off nearly equally for lower and higher values of $h^\star$ independent of $\lambda$, and they decline apparently uniformly with $\lambda$ independent of $h^\star$.    

\begin{figure}
\includegraphics[width=.95\linewidth]{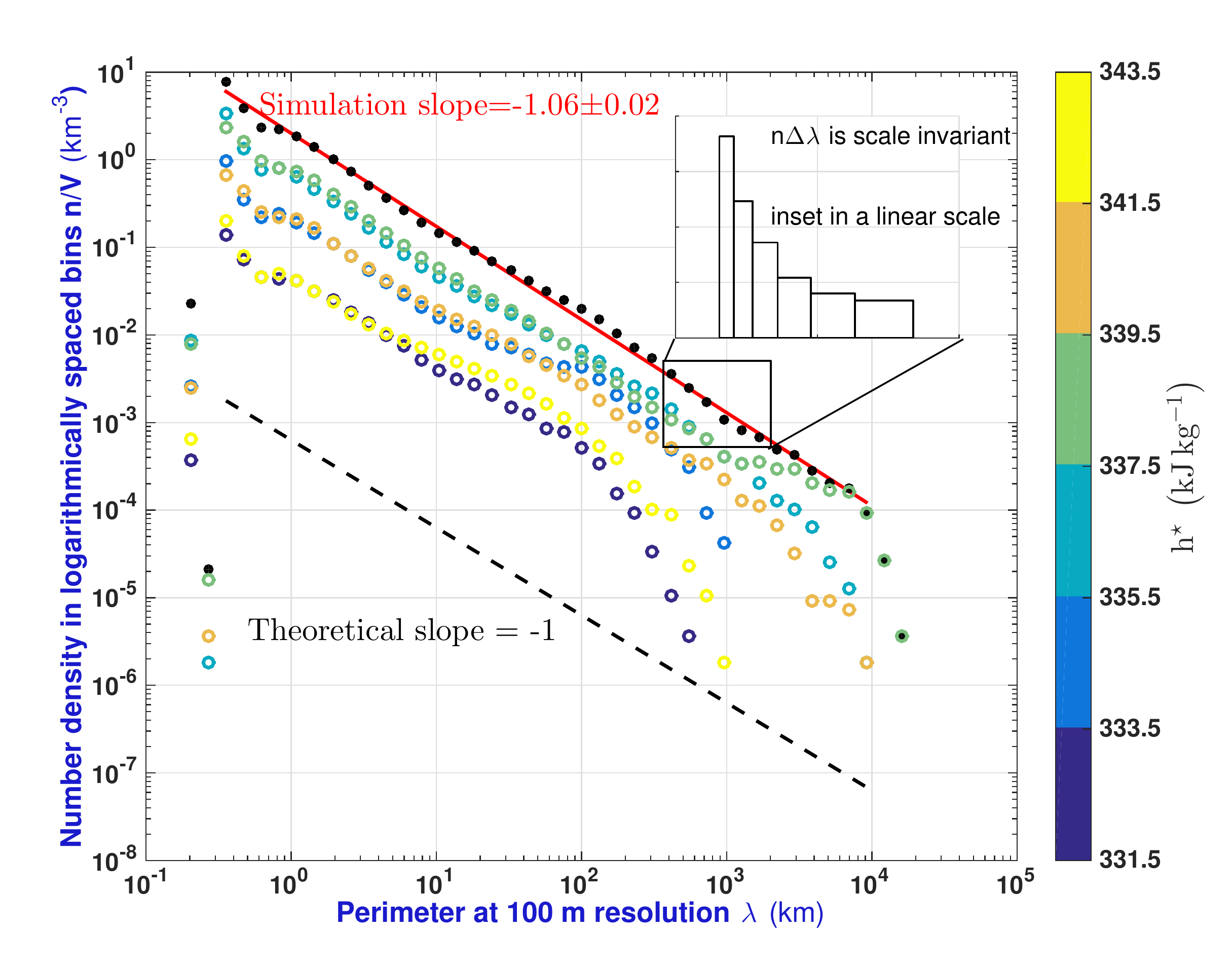}
\caption{\label{fig:distributions}Cloud perimeter density distributions 
evaluated at the intersection of the cloud surfaces and isentropic surfaces and averaged over 12 hours run-time within the Giga-LES
domain, calculated as the number of clouds $n$ in the domain volume $V$ within logarithmically-spaced bins of $\lambda$. Black dots represent cloud perimeter distributions summed over all tropospheric values of the saturated static energy $h^\star$; the red line is the functional fit. Colored open dots are the values calculated at specific moist isentropic
surfaces in $h^\star$ in bands of width $\Delta h = 2$\,kJ\,kg$^{-1}$. The dashed curve represents the power law $n\propto\lambda^{-1}$. The inset illustrates on a linear scale how the product $n\Delta\lambda$ is approximately conserved independent
of $\lambda$ where $\Delta\lambda$ is plotted within logarithmically spaced bins.    }
\end{figure}

Thus, Fig. \ref{fig:3distributions}c suggests that the mathematically well-behaved functional form for cloud frequency distributions described by Eq. \ref{eq:gamma distribution} might indeed be found in the Giga-LES. For the entire ensemble, the number density $n/V$ of clouds with a given perimeter is shown in Fig \ref{fig:distributions}. Note that the ordinate axis is not $n_\lambda/V$ where $n_\lambda = dn/d\lambda$ but instead $n/V$, so any given size bin has a product $n\Delta\lambda/V$ that is proportional to $J$ through Eq. \ref{eq:J_cloud}. Perimeter distributions closely follow the anticipated power law with an exponent of -1 given by Eq. \ref{eq:power law}: on a log-log set of axes, the slope is $-1.06\pm0.01$ spanning over four orders of magnitude in $\lambda$ ranging from 0.4 km to 10,000 km and including 99.6\% of the cumulative cloud perimeter. 

Calculated in equal logarithmically spaced bins $\Delta\ln\lambda$, the product $n\Delta\lambda/V \simeq n\lambda\Delta\ln\lambda/V$ is approximately constant. The implication from Eq. \ref{eq:J_cloud} is that clouds in any given logarithmically spaced perimeter bin contributes as much as any other bin to cloud field circulations across the interface between clouds and clear-skies. 

\begin{figure}
\centering
\includegraphics[width=1\linewidth]{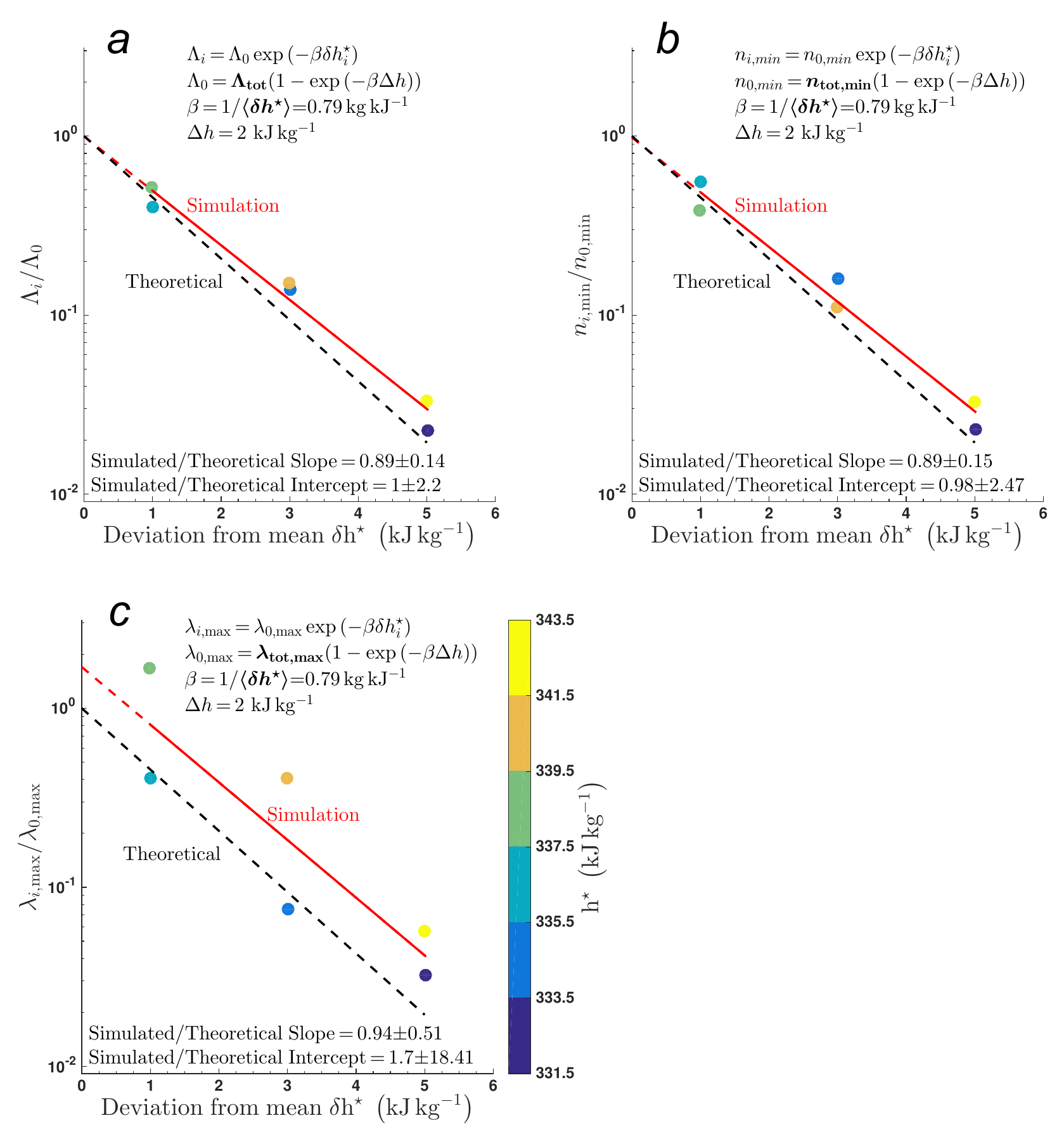}
\caption{\label{fig:Exponentials}Within the Giga-LES, values of  (a) total cloud perimeter, (b) cloud number, and (c) the maximum cloud perimeter as it is defined by where $n\left(\lambda\right)$ falls below 1/e of a power-law extropolation (Fig. \ref{fig:distributions}), shown as a function of the deviation of saturated static energy from a mean value of $h^\star = 337.5$\,kJ\,kg$^{-1}$, and normalized by the respective values for the ensemble. The normalization factor accounts for the bin width $\Delta h$ through e.g., Eq. \ref{eq:Lambda_0} so that the theoretically expected intercept is forced to unity.  Bold text refers to values obtained from the Giga-LES. Theoretically derived functional forms  outlined in Section \ref{sec: across isentropes} are shown by the formulae and the dashed black line. A least squares fit with 95\% uncertainty bounds is shown by the red line.}
\end{figure}

The power law is also evident when the data are divided into  2\,kJ\,kg$^{-1}$ intervals in $h^\star$, although with lower correlation, smaller maximum perimeters $\lambda_{max}$, and lower values of $n_{min}$. As also shown in Fig. \ref{fig:3distributions}c, cloud number densities are lowest for cloud boundary isentropes with the greatest departure $\delta h^\star$ from the mean. 

The functional dependence of these relationships is shown in Fig. \ref{fig:Exponentials}. The total cloud perimeter $\Lambda$ calculated at the model resolution $\xi$ within isentropic layers of width $\Delta h$ centered about $h^\star$, and the number of clouds in the smallest size bin $n_{0,\rm{min}}$, follow a negative exponential with respect to $\delta h^\star$. The results from the simulation are generally consistent with theoretical expectations given by Eqs. \ref{eq:lambda delta h} and \ref{eq:lambda_i}.  The value of $\left<\delta h^\star\right>$ implied by the functional fit is $1.43$\,kJ\,kg$^{-1}$, just 12\% greater than the model value implied by calculation of the square root of the variance in $h^\star$ (Eq. \ref{eq:square root}). Also, the calculated intercept of the fit deviates from the theoretically expected value by <2\%. Incidentally, a similar exponential dependence is shown for the size of the largest clouds $\lambda_{max}$ as a function of $\delta h^\star$, although the formulation for the intercept differs.

To a rough approximation, the Giga-LES results shown in Figs. \ref{fig:3distributions} and \ref{fig:Exponentials} can be seen to apply primarily to high clouds, simply because they dominate the volume. The largest of these clouds are cirrus anvils at the equilibrium height of  $\left<h^\star\right>$. Intuitively, we would expect that cirrus anvils should be largest near this equilibrium level since this is near the location of the level of neutral buoyancy where $\delta h^\star \rightarrow 0$. What is less clear is why clouds should dominate at high levels rather than having similar numbers and sizes as those clouds at lower levels near 2\,km where it also holds that $\delta h^\star \rightarrow 0$ (Fig. \ref{fig:3distributions}). 



For cloud edges with moist static energies within the range $\left<h^\star\right> \pm \left<\delta h^\star\right>$, the average turbulent kinetic energy dissipation rate in the model is $\varepsilon = 3\times10^{-3}$\,m$^2$\,s$^{-3}$. At 100 m horizontal and vertical model resolution $\xi$, this gives a value for the Kolmogorov microscale $\eta = \left(\nu^3/\varepsilon\right)^{1/4}$ where $\nu$ is the kinematic viscosity of air, of 1.1 mm.  For the purposes of diffusion calculations, it follows from  Eq. \ref{eq:D_adjusted}  that $\mathcal{D} \sim 1.3\times10^{-2}$ m$^{2}$ s$^{-1}$.

The total perimeter  of the cloud field is the summation all cloud perimeters at all levels (Eq. \ref{eq:Lambda tot discrete}), where the perimeters are summed in the horizontal and vertical directions using a ``ruler'' with length $\xi$. Normalized by the domain volume, the total cloud perimeter density for $\xi = 100$\,m is $\Lambda_{tot}/V = 59$\,km\,km$^{-3}$. Then, from Eq. \ref{eq:tau_Lambda}, the estimated characteristic time for the mixing of buoyancy across cloud boundaries is $\tau_{cld,clr} = V/\left(\mathcal{D}\left(\xi\right)\Lambda_{tot}\right) \sim146$\,s. For comparison, the mean stability of cloud edges in the model domain is $S = 1.2$\,kJ\,kg$^{-1}$\,km$^{-1}$, so from Eq. \ref{eq:tau_N} this implies an independent but similar estimate for the mixing timescale of $\tau_{stab} = \sqrt{\left<h^\star\right>/Sg} = 166$\,s, consistent with the hypothesis Eq. \ref{eq:timescale equivalence}. Equating these two estimates of $\tau$ (Eq. \ref{eq:Dlambda_S}) the stability $S$ implies a perimeter density $\Lambda_{tot}/V$ equal to 52\,km\,km$^{-3}$ which can be compared to the value of  59\,km\,km$^{-3}$ that is obtained if it is calculated directly from the cloud geometries. 

Thus, there is a difference of just 13\% whether mixing timescales or cloud total perimeters are calculated from the bulk tropospheric stability or from cloud structures. What is raised is that possibility that cloud geometries at steady-state can be seen more fundamentally as an emergent property of bulk atmospheric thermodynamics.

\section*{Discussion and conclusions}
We examined how the statistical distributions of cloud geometries in a tropical cloud field are constrained by the bulk thermodynamic properties of the troposphere. Simple, theoretically derived formulations for equilibrium cloud geometric characteristics closely reproduced the geometry statistics of a spatially and temporally complex dynamic tropical cloud system simulation.   

To achieve this, we replaced deterministic calculations of spatially and temporally evolving state variables in a 3D domain with a time-independent representation of the statistical properties of cloud edge, evaluated in a coordinate system of cloud number, cloud perimeter, and saturated static energy. The motivation for focusing on the interface between clouds and clear air is that total cloud perimeter is a shared property of the two regions; this linear dimension can itself be considered a state variable constraining the cloud field as a whole. 

Across the interface, clouds and clear air compete for flows of air and energy as part of a mixing engine defined by a ``convective potential'' whose energetic magnitude can be determined by either the local stability at cloud edge or the surrounding sub-saturation of clear air. 
At equilibrium, competition for air leads to ensembles of clouds being characterized by mathematically well-defined perimeter distributions. We find that when air parcels along a cloud's perimeter are evaluated within a fixed range of saturated static energy, or within a moist isentropic layer, then both theory and simulations show that cloud number is inversely proportional to perimeter. A few clouds, mostly stratiform cirrus anvils, grow rich in perimeter, while smaller convective clouds are many but poor. The product of the number and the perimeter is constant, meaning that all cloud perimeter classes share equally at mixing air across cloud boundaries. 

Rapid ascent in individual clouds is largely independent of its neighbors, but the ensemble of clouds is not independent of itself. On average, the ascent any one air mass necessarily takes away from the potential for ascent by another. This leads to the result  that cloud edges with the greatest departure of the saturated static energy from the mean are the least frequent, following a negative exponential with respect to the magnitude of the departure.

So, while a power-law or scale invariance is obtained along moist isentropes, a Boltzmann distribution is obtained across moist isentropes. 
Negative exponentials have been described previously for convective mass fluxes \citep{CraigCohen2006} assuming the total mass flux is constrained and the point mass flux $M$ of any particular cloud is independent of its neighbors, i.e. $dn/dM\propto\exp\left(-M/\bar{M}\right)$. Power law behavior has been noted in satellite observations of the linear dimensions of clouds and clear skies \citep{Sengupta1990,Nair1998,NoberGraf2005,WoodField2011,Yuan2011, Yamaguchi2013, Romps2017}, and in the number distribution for the total amount of energy dissipated by tropical cyclones over a range of oceanic basins \citep{Osso2010}. 

Of course, exponential Boltzmann distributions also form a basis for expressing probabilities in quantum phenomena \citep{Andrews1975}, or the size distributions of raindrops \citep{PruppacherKlett1997,Wu2018}, and examples of power-laws over a range of time and spatial scales can be seen throughout nature, in such seemingly dissimilar phenomena as neuronal firing, earthquakes, microbial diversity, war intensity, and personal wealth \citep{buzsaki2004,Newman2005,Locey2016}. Theoretically, power laws, or the property of self-similarity, can also be obtained from a more purely mathematical perspective than was used here, as they emerge when existing objects compete probabilistically for whatever enables their growth in direct proportion to their current size; or, in the size of connected clusters within a lattice if occupancy of any given cell has a predetermined uniform probability \citep{Newman2005}.  

Here, a statistics for cloud macroscopic properties was obtained using more purely thermodynamic reasoning. The advantage of this approach is that it enables a clear link between cloud sizes and atmospheric bulk thermodynamic properties. Integrating over the entire domain, theory and numerical simulations suggest that the total perimeter of all clouds within the atmospheric volume can be linearly related to the square root of the atmospheric stability with respect to a moist adiabat, or inversely with the timescale of buoyancy driven circulations around cloud edges.   

Repeated simulations for varying atmospheric states are required to verify whether total cloud perimeter density does  indeed scale linearly with the buoyancy frequency. However, if the result holds, it would be tantalizing. It would suggest that, statistically speaking, the fine-scale complexity of cloud structures can be tied to the larger scale moist thermodynamic properties of the troposphere. At least within a space of saturated static energy and perimeter, and with the caveat that a basic theory has not been presented for the magnitude of $\left<\delta h^\star\right>$ or $\varepsilon$, an equilibrium cloud field could ``emerge'' from a single point value of stability. 

Admittedly, cloud area is a more familiar and radiatively relevant than cloud perimeter. Here, the observed self-similarity property of clouds allows for use of the fractal relationship $\lambda\propto a^{{D}/{{2}}}$, where $D$ is the fractal dimension $D$ and $a$ is individual cloud cross-sectional area.  For all clouds in the Giga-LES model, a least-squares fit yields $D = 1.38$, which is in approximate agreement with the observational result of $D \sim 1.35$ reported by \citet{lovejoy1982}. 

However, from a radiative standpoint, it is the total cloud area $A$ within the cloudy domain,  summed over a distribution of individual clouds, that is most relevant. To calculate the total cloud area from the total cloud perimeter, we cannot use the same value of $D$ as for individual cloud cross-sections.  Within discrete layers of fixed mean $h^\star$ and width $\Delta h = 2$\,kJ\,K$^{-1}$, a fit to $\Lambda\left(h^\star\right) \propto A\left(h^\star\right) ^{{D_{tot}}/{{2}}}$  yields the result that $D_{tot} = 1.66 \pm 0.14$. This value for the power-law relationship of cloud area to cloud perimeter in the cloud field is relatively invariant to the choice of thickness of the isentropic layer $\Delta h$: if $\Delta h = 0.5$\,kJ\,K$^{-1}$, then $D_{tot} = 1.60 \pm 0.06$. 

This leads to a hypothesis for the role of clouds in climate. Over tropical oceans, satellite measurements indicate that clouds are arranged such that a large negative shortwave cloud radiative forcing is nearly precisely offset by an equally large positive longwave cloud radiative forcing; it has been argued that unless there is a  change in how clouds of different sizes and height are arranged, there should be no expected change in net cloud forcing with climate change, even if total cloud area increases \citep{Hartmann2016}. 

In this context, a recent study used coarse-grid global climate models to argue that higher upper tropospheric stability expected in a warmer climate will cause cirrus anvils to shrink \citep{Bony2016}, although with uncertain net radiative impact.  In contrast, our study suggests that $d\ln\Lambda_{tot}/d\ln S \simeq 1/2$ (Eq. \ref{eq:Dlambda_S}) in which case the aforementioned fractal relation for total cloud cover gives $d\ln A_{tot}/d\ln S \sim 1/D_{tot} \sim  0.6$, that is total cloud area will increase slightly more than than the square root of any change in tropospheric stability that arises from surface warming. 

Because cloud sizes appear to have the property of scale invariance, this relation should be expected to apply equally to the sizes of the largest clouds, which would imply that cirrus anvils will grow rather than shrink. Moreover, if size distributions maintain the invariant mathematical form along and across isentropic layers provided by Eq. \ref{eq:gamma distribution}, it would seem that we should not expect there to be a change in how clouds are arranged with respect to isentropic surfaces: thus, we speculate that climatically induced changes in net cloud radiative impact due to any areal change might be rather small. 

The question of how cloud linear dimensions adjust to changing climate regimes needs  to be explored further using active and passive space-based sensors \citep{Norris2016}, or through sensitivity studies using fine-scale numerical simulations such as those described here.

\acknowledgments{This material is based upon work supported by the National Science Foundation Science and Technology Center for Multi-Scale Modeling of Atmospheric Processes, managed by Colorado State University under cooperative agreement No. ATM-0425247.}

\bibliography{References}
\end{document}